\pdfoutput=1
\documentclass[12pt, a4paper]{article}
\usepackage{jheppub}
\usepackage[usenames, dvipsnames]{xcolor}
\usepackage{graphicx}
\usepackage{microtype}
\usepackage{amssymb} 
\usepackage{amsmath}
\usepackage{framed}
\usepackage{mathtools}
\usepackage{amsfonts}    
\usepackage{dsfont}
\usepackage{pdfpages}
\usepackage{verbatim}
\usepackage{braket}
\usepackage{bm}
\usepackage{tensor}
\usepackage{upgreek}
\usepackage{subcaption}
\usepackage{simpler-wick}
\usepackage{enumitem}
\usepackage[hang, flushmargin]{footmisc}

\usepackage{tikz}
\usetikzlibrary{decorations.markings}
\usetikzlibrary{shapes.misc}
\usetikzlibrary{arrows}
\usetikzlibrary{fit, chains}
\usetikzlibrary{patterns}

\usepackage{listings}
\usepackage{color}

\usepackage[framemethod=default]{mdframed}
\usepackage{etoolbox}

\newenvironment{FramedBox}[1][]{%
	\begin{mdframed}[
		skipabove=7pt,
		skipbelow=7pt,
		rightline=true,
		leftline=true,
		topline=true,
		bottomline=true,
		backgroundcolor=gray!10,
		linecolor=gray!10,
		innerleftmargin=10pt,
		innerrightmargin=10pt,
		innertopmargin=10pt,
		innerbottommargin=10pt,
		leftmargin=0cm,
		rightmargin=0cm,
		linewidth=1pt,
		#1
		]%
		\ignorespaces%
	}{%
	\end{mdframed}%
}

\usepackage[framemethod=default]{mdframed}
\newmdenv[skipabove=7pt,
skipbelow=7pt,
rightline=false,
leftline=false,
topline=false,
bottomline=false,
backgroundcolor=gray!10,
linecolor=gray,
innerleftmargin=5pt,
innerrightmargin=5pt,
innertopmargin=5pt,
innerbottommargin=5pt,
leftmargin=0cm,
rightmargin=0cm,
linewidth=4pt]{eBox}

\usepackage{tikz}
\usetikzlibrary{decorations.pathreplacing, decorations.markings,calc,shapes.misc,decorations.pathmorphing,patterns.meta, math, shadings, backgrounds}

\definecolor{rosewood}{rgb}{0.4, 0.0, 0.04}
\definecolor{pyblue}{RGB}{31, 119, 180}
\colorlet{lightpyblue}{pyblue!30!white}
\definecolor{pyorange}{RGB}{255, 127, 14}
\colorlet{lightpyorange}{pyorange!30!white}
\definecolor{pygreen}{RGB}{44, 160, 44}
\colorlet{lightpygreen}{pygreen!30!white}
\definecolor{pyred}{RGB}{214, 39, 40}
\colorlet{lightpyred}{pyred!30!white}
\definecolor{lightgray}{gray}{0.9}

\usepackage{hyperref}
\hypersetup{
	pdfencoding=unicode,
	colorlinks=true,
	urlcolor=rosewood,
	linkcolor=pyblue,
	citecolor=pygreen,
	pdftitle={Massive Cosmological Correlators from Flat Space: a Laplace-Space Approach},
	pdfauthor={Nathan Belrhali, Arthur Poisson, Sebastien Renaux-Petel},
	pdfdisplaydoctitle=true,
	pdfstartview=FitH,
	linktocpage=true
}

\def \Im {\mathrm{Im}}
\def \Re {\mathrm{Re}}
\def \Disc {\text{Disc}}

\def \d {\mathrm{d}}

\def \k {\boldsymbol{k}}
\def \x {\boldsymbol{x}}

\def \F {\mathcal{F}}

\def\lambdat{\tilde{\lambda}}

\newcommand{\ba}{\begin{aligned}}
	\newcommand{\ea}{\end{aligned}}

\def\tk{\tilde{\kappa}}

\allowdisplaybreaks[1]
\title{Massive Cosmological Correlators from Flat Space: a Laplace-Space Approach}

\author{Nathan Belrhali,}
\author{Arthur Poisson,}
\author{S\'ebastien Renaux-Petel}

\affiliation{Institut d'Astrophysique de Paris, CNRS, Sorbonne Universit\'e,
98 bis bd Arago, 75014 Paris, France}

\abstract{We develop a new approach to cosmological correlators, built on a simple
	physical fact: deep inside the Hubble radius every mode oscillates as a flat-space plane wave,
	the curvature of spacetime making itself felt only as the mode is stretched towards the horizon.
	A Laplace transform turns this observation into a computational tool, resolving each curved-space
	mode function into a continuous superposition of plane waves labelled by a dual
	variable and dressed by a kernel that encodes the spacetime geometry, field content and dynamics. Every time integral then reduces to an elementary flat-space one, yielding simple diagrammatic rules for cosmological correlators.
	We illustrate the construction on the massive single-exchange correlator. The Laplace
	representation makes its total- and partial-energy singularities transparent ``from flat
	space'', and yields a single closed-form, rapidly convergent series valid throughout the entire
	kinematic domain. Although developed for conformally coupled fields exchanging massive scalars in de
	Sitter, the approach carries over essentially unchanged to virtually all situations of interest
	in primordial cosmology.}

\begin{document}
	
	\setcounter{tocdepth}{3}
	\maketitle
	\setcounter{page}{2}
	
	\newpage
	\section{Introduction}
	
	Locally, any spacetime looks like flat space. During inflation this equivalence principle has a
	sharp dynamical counterpart: a mode deep inside the Hubble radius does not yet feel the expansion
	of spacetime and oscillates as a plane wave, the geometry making itself felt only as the mode is
	stretched towards the horizon. In this paper we turn this observation into a computational tool.
	We show that the time-dependent building blocks of cosmological correlators admit a simple
	integral representation over plane waves, and that cosmological correlators are
	thereby related to their flat-space counterparts by an integral transform whose kernel encodes
	the spacetime geometry together with the field content and dynamics of the system.
	
	Cosmological correlators---the late-time $n$-point functions of the primordial curvature and
	tensor perturbations---are the observables of inflation, which hold the promise to uncover primordial physics at very high energies that is inaccessible otherwise \cite{Achucarro:2022qrl}. Despite their importance, computing and understanding them remains a challenge, due to the breaking of time-translation invariance caused by the expansion of the universe. This has two consequences: mode functions are distorted from plane waves to special functions, Hankel functions for a massive fields and Whittaker functions once a chemical
	potential is switched on for a spin-1 field; and correlators are given by nested time integrals over products of such mode functions. 
    A wealth of techniques has been developed to tackle this issue: the
	cosmological bootstrap, which uses the boundary differential equations obeyed by correlators together with their singularity structure, both in its original de Sitter-invariant
	formulation and in its boostless extensions~\cite{Arkani-Hamed:2018kmz,Baumann:2019oyu,Baumann:2020dch,Pajer:2020wxk,Pimentel:2022fsc,Jazayeri:2022kjy,Wang:2022eop,Qin:2022fbv,Qin:2023ejc};
	Mellin-space representations of the bulk time integrals~\cite{Sleight:2019hfp,Sleight:2019mgd,Sleight:2020obc,Sleight:2021plv,Qin:2022fbv,Qin:2023bjk}; dispersive, unitarity- and
	analyticity-based methods reconstructing correlators from their discontinuities~\cite{Goodhew:2020hob,Jazayeri:2021fvk,Melville:2021lst,Goodhew:2021oqg,Cabass:2021fnw,DiPietro:2021sjt,Meltzer:2021zin,Stefanyszyn:2023qov,Qin:2023bjk,Liu:2024xyi,Liu:2026jzn};
the
	K\"all\'en-Lehmann representation in de Sitter, which trades a loop for a continuous sum of
	tree-level exchanges~\cite{Hogervorst:2021uvp,DiPietro:2021sjt,Xianyu:2022jwk,Loparco:2023rug,Loparco:2025azm};
    family-tree decompositions yielding closed-form expressions for nested time
	integrals~\cite{Xianyu:2022jwk,Qin:2023bjk,Qin:2023ejc,Xianyu:2023ytd,Liu:2024str,Fan:2024iek,Fan:2025scu}; 
    the kinematic-flow~\cite{Arkani-Hamed:2023kig,Arkani-Hamed:2023bsv,De:2023xue,Baumann:2024mvm,Westerdijk:2026msm,Ke:2026laa,Baumann:2026atn};
    spectral representation in the de Sitter frequency-momentum space (Kontorovich-Lebedev-Fourier)~\cite{Melville:2024ove,Werth:2024mjg,Nowinski:2025cvw,Lee:2025kgs,Belrhali:2026ktb,Belrhali:2026rkn,Grafe:2026qsm}; the Grassmannian formulation~\cite{Arundine:2026fbr,De:2026shn,Huang:2026tsh,Arundine:2026myr};
	or direct numerical integration of the cosmological flow of the correlators~\cite{Werth:2023pfl,Pinol:2023oux,Jazayeri:2023xcj,Werth:2024aui}.

In this work accompanying the Letter \cite{Belrhali:2026laplace-letter}, we develop a new approach, built on the early-time behaviour of the modes. The key step is a Laplace transform: after the Wick rotation that turns the early-time oscillation $e^{-ik\tau}$ into a decaying exponential, each bulk function, viewed as a function of $z=k\tau$, is Laplace-transformed in $z$. Inverting the transform reconstructs the (Wick-rotated) mode, a generic operation valid for essentially any bulk function.\footnote{A complementary recent approach also starts from an integral representation of the massive mode functions, using it to derive the differential equations obeyed by the corresponding wavefunction coefficients~\cite{Baumann:2026atn}.} What turns this into a computational tool is the Bunch-Davies behaviour of the modes at early times, which fixes the analytic structure of the transform: analytic for $\Re(\lambda)>-1$, it has a branch point at $\lambda=-1$, the dual image of that early-time behaviour. Deforming the inversion contour onto the corresponding cut and rotating back to real time recasts the mode function as a continuous superposition of plane waves $e^{-i\lambda k\tau}$, labelled by the dual variable $\lambda$ and weighted by the discontinuity across the cut, the kernel that dresses the flat-space content back into the curved-space one.
The same structure can be reached from a complementary direction: starting from a plane wave behaviour, the mode function
	is corrected by an asymptotic series as one departs from the infinite past; that series is divergent but Borel summable, and
	resumming it reconstructs the exact mode function, returning the very same plane-wave representation in the de Sitter-invariant case while differing from it beyond that case (Appendix~\ref{sec:Borel-Whittaker}).
	
	With this representation in hand, the time integrals of the in-in formalism collapse. In the
	dual Laplace space each massive line becomes a plane wave, so every bulk time integral reduces
	to an elementary flat-space (massless) one, while the geometry, field content and dynamics
	survive entirely in the known $\lambda$-space kernel. We turn this into a
	set of diagrammatic rules that produce, directly from a diagram with polynomial interactions, the Laplace-space integrand for de
	Sitter correlators with conformally coupled external legs and massive internal exchanges, valid equally at tree and loop level. In an appendix, we also provide the Laplace representation in the more general case of a massive spin-1 field with a helical chemical potential dressed by an arbitrary twist, i.e.\ a
	twisted Whittaker mode function.

    We apply our formalism to
 the massive single-exchange, which demonstrates its two benefits. On the conceptual side, the Laplace representation makes the analytic structure of the
	correlator transparent ``from flat space'': the total- and partial-energy singularities emerge
	directly from the flat-space data, the former as a pinch at the early-time corner of the dual
	integration domain and the latter as a flat-space pole reaching its boundary. On the practical side,
	the same representation yields what is, to our knowledge, the most powerful analytic handle on this
	paradigmatic correlator: a single closed-form and very rapidly convergent series,
	valid throughout the entire kinematic domain with no patching of separate expansions.

	Although we develop everything in detail for a definite setup---polynomial interactions of a conformally
	coupled field with massive scalars in de Sitter---the construction is far more general. Trading every
	non-trivial time dependence for plane waves through a Laplace representation carries over essentially
	unchanged to wavefunction
	coefficients, de Sitter-breaking setups with derivative interactions, non-trivial sound speeds, 
	spin-1 fields with a chemical potential and time-dependent couplings. We collect these extensions in Section~\ref{sec:generalisations}.

	Several groups
	have shown how cosmological correlators can be obtained by dressing flat-space amplitudes~\cite{Chowdhury:2023arc,Chowdhury:2025ohm,Chowdhury:2026upp,Das:2025qsh,Das:2026vfv,Ansari:2026xkm,Ansari:2026sjf}. In practice, however, these constructions
	have been confined to theories of massless or conformally coupled fields and have set aside the exchange of massive fields, which is singularly more involved. By contrast, our Laplace approach handles this case and essentially any theory of interest in primordial cosmology. 
	
	Finally, it is instructive to set this approach against the other integral transforms used in the field,
	since each rests on a different symmetry. Methods based on the Mellin transform rely on the
	late-time conformal symmetry of the boundary; de Sitter momentum-space Kontorovich-Lebedev-Fourier methods exploit the full de Sitter invariance together with spatial translations. The Laplace
	transform, by contrast, is rooted in
	an \emph{asymptotic} early-time symmetry---the flat-space behaviour that the modes recover deep inside
	the horizon---which is exactly why a Borel resummation of the early-time asymptotic series
	reconstructs the exact mode function.

	The paper is organised as follows. Section~\ref{sec: Laplace space} develops the method: the
	Laplace transform of bulk time integrals and its analytic structure, the dual equation that
	determines the mode function and selects it through its early-time behaviour, the resulting
	plane-wave representation, and its Borel-resummation counterpart. Section~\ref{sec: correlators}
	recalls the in-in formalism, derives the Laplace-space diagrammatic rules, revisits the
	single-exchange correlator carrying it through to a closed form, and lays out the generalisations of the
	method. Appendix~\ref{app: special functions} collects special-function material.
	Appendix~\ref{app:Whittaker} treats the general twisted-Whittaker case by both the Laplace and the
	Borel routes. Appendix~\ref{app: total energy series} re-derives the total- and partial-energy
	singularities of the single-exchange correlator from its master series. Appendix~\ref{app:WFU} gives the corresponding diagrammatic rules for wavefunction coefficients.
	
	\section{Curved spacetime dynamics from plane waves}
	\label{sec: Laplace space}
	
	\subsection{Time integrals as Laplace transforms}
	\paragraph{Master integral.} In the perturbative evaluation of cosmological correlators, one encounters multiple, nested conformal-time integrals of products of mode functions. The elementary object underlying them is the time integral of a single mode function against a plane wave, which we take as our starting point:
	\begin{equation}\label{eq: Fhat physical def}
		\hat{\F}\left(\lambda\right) = \int\displaylimits_{-\infty(1+ i \epsilon)}^0\d\tau \;k_Ie^{- i k_E \tau}\F(k_I\tau) = \int\displaylimits_{-\infty(1+ i \epsilon)}^0\d z e^{- i \lambda z}\F(z)\;,
	\end{equation}
	where $k_{I,E}$ denote two different momenta, we defined $\lambda=k_E/k_I$, and the $i\epsilon$ prescription consistently prepares the Bunch-Davies vacuum state at early times by enforcing the convergence of the integral. What follows holds for any bulk function with the appropriate asymptotic behaviour but, for simplicity, we will call $\F$ a mode function, as we will indeed apply our formalism to (suitably rescaled) mode functions $\F$ encountered in inflationary cosmology. 
	Considering the Bunch-Davies state, the latter behave like a plane wave in the far past (we neglect any subleading dependence here):
	\begin{equation}
		\F(z)\underset{z\to-\infty}{\sim} e^{-i z}\;,
	\end{equation}
	which is the only requirement.
	Therefore, since the integrand in~\eqref{eq: Fhat physical def} scales like $e^{-i(\lambda+1)z}$, the integral is convergent in the region $\Re(\lambda)>-1$ which includes all the possible physical kinematic configurations, and other areas of the complex $\lambda$ plane can be accessed after analytic continuation of $\hat{\F}(\lambda)$. 
	\paragraph{Wick rotation and inversion.}
	The tilted axis over which we need to integrate appears to be inconvenient for the practical evaluation of~\eqref{eq: Fhat physical def}. This can be addressed by deforming the integration contour to the imaginary axis:
	\begin{equation}\label{eq: Master in-in integral Wick}
		\hat{\F}(\lambda) =  i\int\displaylimits^\infty_0\d z e^{-\lambda z}\F(- i z) + \textrm{Residues}\;.
	\end{equation}
	Contributions from residues arise if the function $\F(z)$ has poles in the lower left quadrant. 
	This is not the case in the situations we discuss in this paper, so we drop this term in what follows, but even if present, our representation of bulk functions as superpositions of plane waves goes through unaffected. The integral~\eqref{eq: Master in-in integral Wick} can be identified as the Laplace transform of the function $\F(- i z)$ as $\hat{\F}(\lambda) = i \mathcal{L}\left[\F(- i z)\right](\lambda)$. As such, it can be inverted to give:
	\begin{equation}\label{eq: inverse Laplace tranform def}
		i\F(-iz) =\mathcal{L}^{-1}\left[\hat{\F}(\lambda)\right]( z)\equiv \int\displaylimits_{c-i\infty}^{c+i\infty}\frac{\d \lambda}{2\pi i}e^{\lambda z}\hat{\F}(\lambda)\;,
	\end{equation}
	where $c$ is an arbitrary constant such that the Laplace transform is analytic in the half plane $\Re(\lambda)>c$. As the integral converges towards finite values for $\Re(\lambda)>-1$, one can always take $c$ to be in this region of the complex plane. 
	\paragraph{Analytic structure and the cut.}  In the next section, to turn the inversion~\eqref{eq: inverse Laplace tranform def} into a useful representation, we will
	close its contour to the left (see Fig.~\ref{fig: contour deformation} below), which requires continuing $\hat{\F}$ beyond the strip
	$\Re(\lambda)>-1$ where it was defined. Deforming the contour of~\eqref{eq: Master in-in integral Wick}
	extends $\hat{\F}$ to an analytic function on the whole $\lambda$ plane minus a branch cut along
	$\lambda\in(-\infty,-1)$. Two properties of this continued transform matter
	below, both fixed by the mode function itself. First, its asymptotic behaviours: as $\lambda\to\infty$ one has $\hat{\F}(\lambda)\sim\lambda^{-\Delta^0_\F-1}$, set by the
	small-$z$ exponent $\F(z)\sim z^{\Delta^0_\F}$, while as $\lambda\to-1$ one has
	$\hat{\F}(\lambda)\sim(1+\lambda)^{-\Delta^\infty_\F-1}$, set by the early-time exponent in
	$\F(z)\sim z^{\Delta^\infty_\F}e^{-iz}$; the latter is what makes $\lambda=-1$ a branch
	point, which we call the (Laplace-space) Bunch-Davies singularity.\footnote{The marginal case $\Delta^\infty_\F=-1$, realised by the massive mode function
		below, degenerates: the power turns into a logarithm. We treat it in
		Section~\ref{subsec: dual equation}.} Second, the discontinuity across the cut is inherited from
	the time domain---when $\F(-iz)$ is cut along the negative real axis, $\hat{\F}$ is cut on
	$(-\infty,-1)$ with
	\begin{equation}\label{eq: disc reflected}
		\Disc_\lambda\big[\hat{\F}(\lambda)\big] = \mathcal{L}\big[-i\,\Disc_z[\F(iz)]\big](-\lambda)\;.
	\end{equation}
	Both statements are established in the inset below.
	
	\begin{FramedBox}
		\noindent\textbf{Inset: analytic structure of $\hat{\F}$.}\;
		Deforming the contour of~\eqref{eq: Master in-in integral Wick} continues the Laplace transform to
		any $\lambda\in\mathbb{C}\setminus\{-1\}$, reproducing the defining integral for $\Re(\lambda)>-1$ (see, e.g.~\cite{Liu:2024xyi} for similar analysis):
		\begin{equation}
			\hat{\F}(\lambda) = i\int\displaylimits_0^{\frac{1}{\lambda+1}\infty}\d z\; e^{-\lambda z}\,\F(-iz)\;.
		\end{equation}
		The endpoint behaviours follow by scaling estimates. As $\lambda\to\infty$ the exponential confines
		the integral to $z\to0$, where $\F(z)\sim z^{\Delta^0_\F}$, giving
		\begin{equation}\label{eq: Laplace general large lambda}
			\hat{\F}(\lambda)\underset{\lambda\to\infty}{\sim}\;
			i\int\displaylimits_0^{\infty}\d z\,(-iz)^{\Delta^0_\F}e^{-\lambda z}
			\;\sim\; \frac{1}{\lambda^{\Delta^0_\F+1}}\;.
		\end{equation}
		Near $\lambda=-1$, rescaling $z=s/(\lambda+1)$ and using the early-time form
		$\F(z)\sim z^{\Delta^\infty_\F}e^{-iz}$ gives
		\begin{equation}\label{eq: leading behaviour Laplace -1}
			\hat{\F}(\lambda)\underset{\lambda\to-1}{\sim}\;
			\frac{i}{\lambda+1}\int\displaylimits_0^{\infty}\d s\; e^{-s}
			\left(\frac{-is}{\lambda+1}\right)^{\Delta^\infty_\F}
			\;\sim\; \frac{1}{(\lambda+1)^{\Delta^\infty_\F+1}}\;.
		\end{equation}
		Finally, the discontinuity is obtained from the same continued contour. The two terms in
		\begin{equation}
			\begin{aligned}
				\Disc_\lambda\big[\hat{\F}(\lambda)\big] &\equiv \hat{\F}(\lambda+i\epsilon)-\hat{\F}(\lambda-i\epsilon) \\
				&= i\int\displaylimits_0^{\frac{1}{\lambda+1+i\epsilon}\infty}\d z\; e^{-\lambda z}\F(-iz)
				-i\int\displaylimits_0^{\frac{1}{\lambda+1-i\epsilon}\infty}\d z\; e^{-\lambda z}\F(-iz)
			\end{aligned}
		\end{equation}
		differ only by the tilt of their contour at infinity. Rescaling $z\to(\lambda+1)z$ brings both onto
		a common ray, so that the $\pm i\epsilon$ difference is carried entirely by $\F$ and amounts to
		its own discontinuity. Back in the variable $z$, one gets:
		\begin{equation}
			\Disc_\lambda\big[\hat{\F}(\lambda)\big]
			= -i\int\displaylimits_0^{\frac{1}{1+\lambda}\infty}\d z\; e^{-\lambda z}\,\Disc_z\big[\F(-iz)\big]\;,
		\end{equation}
		which, for a cut of $\F(-iz)$ along the negative real axis, reduces to~\eqref{eq: disc reflected}.
	\end{FramedBox}
	
	In practice we do not evaluate the right-hand side of~\eqref{eq: disc reflected}. Once
	$\hat{\F}(\lambda)$ is known in closed form---which the dual equation of the next subsection
	provides---its discontinuity is read off directly in the $\lambda$ plane from the connection
	formulae of the relevant special functions. This is the route followed both in the example below
	and in the more general cases of Appendix~\ref{app:Whittaker}; the inversion itself is carried out
	once the closed form is in hand (see \emph{Inversion} below).
	
	\subsection{Plane-wave representation of mode functions}\label{subsec: dual equation}
	
	\paragraph{From the time evolution to a dual equation.}
	The construction so far represents the mode function through its Laplace transform
	$\hat{\F}$, but does not yet say how to compute the latter. A direct route is available
	whenever $\F$ obeys a linear differential equation in time, as is always the case for a free
	field. The Laplace transform trades multiplication by $z$ for differentiation in $\lambda$,
	and differentiation in $z$ for multiplication by $\lambda$ up to boundary terms,
	\begin{equation}\label{eq: transform rules}
		\mathcal{L}[z\,\psi](\lambda) = -\partial_\lambda \mathcal{L}[\psi](\lambda)\,,\qquad
		\mathcal{L}[\psi'](\lambda) = \lambda\,\mathcal{L}[\psi](\lambda) - \psi(0)\,.
	\end{equation}
	For an equation of motion with constant coefficients this would turn the differential
	equation into an \emph{algebraic} one. This is not our situation: the equations of motion in
	a cosmological background carry explicit powers of $z=k\tau$, so the dual equation is itself
	a differential equation in $\lambda$.\footnote{Equivalently, written with inverse powers of
		$z$ (such as $z^{-1}\partial_z$ or $z^{-2}$), the equation of motion maps these to integral
		operators $\int_\lambda^\infty\!\d\lambda'$ in Laplace space, i.e.\ to an integro-differential
		equation; differentiating in $\lambda$ enough times removes the integrals and returns the
		same differential equation. It is most economical to first multiply the equation of motion
		by the powers of $z$ needed to render all coefficients polynomial.} The boundary term is
	generated only at the endpoint of the time integral: at early times the Bunch-Davies
	fall-off kills the contribution from $z\to\infty$, leaving only the late-time endpoint
	$z\to0$, whose role we return to below.
	
	\paragraph{The dual equation for the massive mode function.}
	We make this explicit on the function relevant for a massive scalar field in de Sitter. The
	physical mode function is built from the Hankel function $H^{(1)}_{i\mu}(-k\tau)$ and the object
	of interest in this context is
	\begin{equation}
		\F(z) = \frac{H^{(1)}_{i\mu}(-z)}{\sqrt{-z}}\,.
		\label{eq:F-Hankel}
	\end{equation}
	The index $i\mu$ carries the mass: $\mu$ is real for heavy fields (the principal series),
	and purely imaginary for light fields (the complementary series), with $0\le|\Im(\mu)|\le\tfrac32$,
	the massless case sitting at $|\Im(\mu)|=\tfrac32$. We build the transform first in the regime
	where the defining integral~\eqref{eq: Fhat physical def} converges---with $\F(z)\sim z^{-1/2\pm i\mu}$ near 0, this imposes $|\Im(\mu)|<\tfrac12$---and analytically continue the resulting closed form in $\mu$ afterwards. As a result, the final expression \eqref{eq: massive mode plane wave} below will be valid in the whole physical range.

	From
	Bessel's equation, $\F$ obeys
	\begin{equation}
		z^2\,\F'' + 2z\,\F' + \left(z^2 + \mu^2 + \tfrac14\right)\F = 0\,.
	\end{equation}
	The Wick rotation that turns the early-time plane wave into a decaying exponential,
	$h(z)\equiv\F(-iz)$, flips the sign of the $z^2$ term,
	\begin{equation}\label{eq: Wick rotated massive}
		z^2\,h'' + 2z\,h' + \left(\mu^2 + \tfrac14 - z^2\right)h = 0\,.
	\end{equation}
	We now apply the transform rules~\eqref{eq: transform rules} to the Wick-rotated
	equation~\eqref{eq: Wick rotated massive} term by term. The transform of $h$ enters through
	$\mathcal{L}[h]$, related to the physical object by $\hat{\F}=i\,\mathcal{L}[h]$, so that
	both obey the same homogeneous equation. Furthermore, the boundary terms coming from the Laplace transform are formally divergent, since
	$h(z)\sim z^{-1/2\pm i\mu}$ as $z\to0$, but cancel in the combinations $\mathcal{L}[z^2h'']=\partial_\lambda^2\!\big(\lambda^2\mathcal{L}[h]-\lambda h(0)-h'(0)\big)$ and $\mathcal{L}[zh']=-2\partial_\lambda\!\big(\lambda\mathcal{L}[h]-h(0)\big)$. The dual equation is
	therefore homogeneous and reads:
	\begin{equation}\label{eq: Legendre dual equation}
		(\lambda^2-1)\,\hat{\F}''(\lambda) + 2\lambda\,\hat{\F}'(\lambda)
			+ \left(\mu^2+\tfrac14\right)\hat{\F}(\lambda) = 0
	\end{equation}
	which is the Legendre equation of degree $i\mu-\tfrac12$. The time evolution of the massive field has thus been traded for
	the Legendre equation in the Laplace variable $\lambda$. This term-by-term transform is carried out within the convergence
	strip $\Re(\lambda)>-1$; the dual equation, and the solution it selects, are then continued
	analytically to the rest of the $\lambda$ plane, where the closed form below acquires its branch
	cut.
	
\paragraph{Selecting and normalising the solution from Bunch-Davies.}
Equation~\eqref{eq: Legendre dual equation} admits the two Legendre solutions
$P_{i\mu-1/2}$ and $Q_{i\mu-1/2}$, which behave differently at the two singular points $\lambda=\pm 1$ of the dual equation \eqref{eq: Legendre dual equation}; $P_{i\mu-1/2}$ is regular at $\lambda=1$ and carries a logarithmic branch at $\lambda=-1$, while $Q_{i\mu-1/2}$ is finite at $\lambda=-1$ and carries a logarithmic branch at $\lambda=1$. This enables one to select and normalise the physical solution:
as $\hat{\F}$ is analytic at $\lambda=+1$, the solution is simply proportional to $P_{i\mu-1/2}$. In other words, the singularity of $Q$ at $\lambda=+1$ is the dual image of the existence of negative frequency modes $e^{iz}$ in the asymptotic past. We call this Laplace-space singularity the excited state singularity, which is here excluded from the outset.~\footnote{Mathematically, one could equally use how the late-time behaviour $z \to 0$ manifests in Laplace space for $\lambda \to \infty$, but the early-time reasoning applies more generally.} The normalisation is then fixed at $\lambda=-1$, where the early-time power of the mode function is the marginal value $\Delta^\infty_\F=-1$ mentioned below \eqref{eq: leading behaviour Laplace -1}: the naive power-law estimate $(\lambda+1)^{-\Delta^\infty_\F-1}$ degenerates and, pushed one order further, becomes a logarithm. Matching its coefficient against the behaviour $P_{i\mu-1/2}(\lambda)\;\underset{-1}{\sim}\;-\frac{\cosh(\pi\mu)}{\pi}
    \ln(1+\lambda)$ fixes the $\mu$-dependent normalisation of $P$:~\footnote{The coefficient follows from pushing
the estimate \eqref{eq: leading behaviour Laplace -1} to the marginal case
$\Delta^\infty_\F=-1$. After the Wick rotation the early-time tail reads
$\F(-iz)\sim -iC\,z^{-1}e^{-z}$ with the Bunch-Davies amplitude
$C=\sqrt{2/\pi}\,e^{\frac{\pi\mu}{2}-\frac{i\pi}{4}}$, fixed by the large-argument Hankel
asymptotics. For $\Delta^\infty_\F\neq-1$ the substitution $z=s/(1+\lambda)$ produces the
power $(1+\lambda)^{-\Delta^\infty_\F-1}$; at $\Delta^\infty_\F=-1$ this degenerates to a
constant and the leading singularity comes from the large-$z$ end of the integral,
$\int^{\infty}\!\tfrac{\d z}{z}\,e^{-(1+\lambda)z}\sim-\ln(1+\lambda)$, so that
$\hat{\F}(\lambda)\sim -C\ln(1+\lambda)$ near $\lambda=-1$.}
\begin{equation}\label{eq: Laplace transform massive closed}
    \hat{\F}(\lambda) = \sqrt{2\pi}\,e^{\frac{\pi\mu}{2}+\frac{i\pi}{4}}\,
    \frac{P_{i\mu-1/2}(\lambda)}{i\cosh(\pi\mu)}\,.
\end{equation}
This reasoning is the Laplace-space avatar of the statement that Bunch-Davies initial
conditions fully determine the mode function.

	
	\begin{figure}[t]
		\centering
		\begin{tikzpicture}[
			axisline/.style={gray!55, ->, thin},
			cut/.style={rosewood, very thick, decorate,
				decoration={zigzag, segment length=4.2pt, amplitude=1.3pt}},
			cont/.style={pyblue, very thick},
			bigarc/.style={pyblue!55, dashed, thick},
			midarrow/.style={postaction={decorate},
				decoration={markings, mark=at position #1 with
					{\fill[pyblue] (-1.3pt,-2.3pt) -- (2.6pt,0pt) -- (-1.3pt,2.3pt) -- cycle;}}}]
			
			\begin{scope}
				\draw[axisline] (-4.1,0) -- (2.4,0) node[below right=-1pt]{$\operatorname{Re}\lambda$};
				\draw[axisline] (0,-2.7) -- (0,2.7) node[right]{$\operatorname{Im}\lambda$};
				\draw[cut] (-1,0) -- (-4.0,0);
				\node[rosewood] at (-2.55,0.42) {\small cut $(-\infty,-1)$};
				\filldraw[rosewood] (-1,0) circle (2.1pt) node[below=3pt]{$-1$};
				\draw[cont, midarrow=0.62] (0.8,-2.45) -- (0.8,2.45);
				\node[pyblue] at (1.18,2.25) {$\gamma$};
				\draw[gray!70] (0.8,0.09) -- (0.8,-0.09);
				\node[below=1pt] at (1.02,-0.05) {\small $c$};
				\node[align=center] at (-0.8,-3.25)
				{\small (a) Bromwich inversion contour~\eqref{eq: inverse Laplace tranform def}};
			\end{scope}
			
			\begin{scope}[xshift=8.6cm]
				\draw[axisline] (-4.1,0) -- (2.4,0) node[below right=-1pt]{$\operatorname{Re}\lambda$};
				\draw[axisline] (0,-2.7) -- (0,2.7) node[right]{$\operatorname{Im}\lambda$};
				\draw[cut] (-1,0) -- (-4.0,0);
				\draw[bigarc] (0.8,2.45)  to[out=158,in=74]  (-4.0,0.32);
				\draw[bigarc] (0.8,-2.45) to[out=202,in=-74] (-4.0,-0.32);
				\node[pyblue!60, align=center] at (1.95,1.65) {\small arcs at $\infty$\\[-1pt]\small (dropped)};
				\draw[cont, midarrow=0.55] (-4.0,-0.32) -- (-1,-0.32);
				\draw[cont, midarrow=0.5]  (-1,-0.32) arc[start angle=-90, end angle=90, radius=0.32];
				\draw[cont, midarrow=0.55] (-1,0.32) -- (-4.0,0.32);
				\node[pyblue] at (-2.95,0.62)  {\small $\lambda+i\epsilon$};
				\node[pyblue] at (-2.95,-0.62) {\small $\lambda-i\epsilon$};
				\draw[gray!70, |-|] (-1,0) -- ($(-1,0)+(0:0.32)$);
				\node[pyblue] at (-0.32,0.2) {\small $C_\epsilon$};
				\filldraw[rosewood] (-1,0) circle (2.1pt);
				\node[rosewood] at (-1.32,-0.62) {$-1$};
				\node[align=center] at (-0.05,-1.55)
				{\small $C_\epsilon \displaystyle\sim\epsilon^{-\Delta^\infty_\F}\xrightarrow[\epsilon\to0]{}0$};
				\node[align=center] at (-0.8,-3.25)
				{\small (b) deformed contour wrapping the cut~\eqref{eq: dispersion Wick rotated}};
			\end{scope}
			
		\end{tikzpicture}
		\caption{Contour deformation in the complex $\lambda$ plane used to pass from the
			Laplace inversion~\eqref{eq: inverse Laplace tranform def} to the dispersive
			representation~\eqref{eq: dispersion Wick rotated}.
			\textbf{(a)} The Bromwich contour $\gamma$ runs vertically at
			$\operatorname{Re}\lambda=c>-1$, to the right of the Bunch-Davies branch point $\lambda=-1$ and of the
			cut along $(-\infty,-1)$.
			\textbf{(b)} Closing $\gamma$ to the left and dropping the arcs at infinity (dashed),
			the contour collapses onto a path that hugs the cut, shown in plain blue.}
		\label{fig: contour deformation}
	\end{figure}
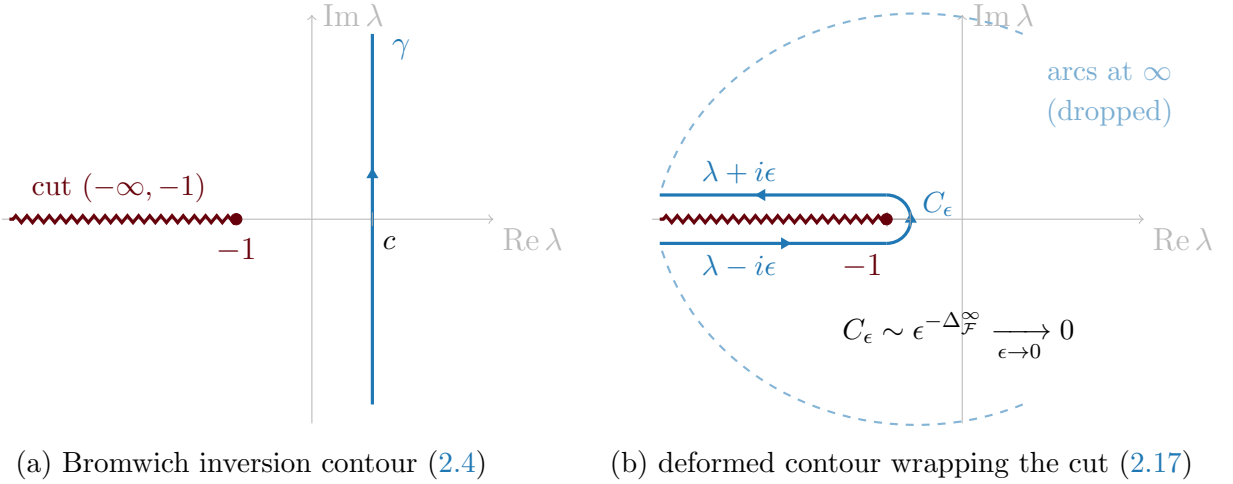

	\paragraph{Inversion.}
	With the closed form \eqref{eq: Laplace transform massive closed} at hand, we recover the mode
	function by inverting the Laplace transform. We close the contour on the left, as shown in Fig.~\ref{fig: contour deformation}. The arcs at infinity do not contribute, and we are left with the contour in plain blue: a segment just below the cut
	($\lambda-i\epsilon$), a small semicircle $C_\epsilon$ of radius $\epsilon$ to the right of
	$\lambda=-1$, and a segment just above the cut ($\lambda+i\epsilon$).
	Given the endpoint
	behaviour~\eqref{eq: leading behaviour Laplace -1},
	$\hat{\F}(\lambda)\sim(1+\lambda)^{-\Delta^\infty_\F-1}$, the $C_\epsilon$ contribution
	scales as $\epsilon^{-\Delta^\infty_\F}$ (degenerating to $\epsilon\ln\epsilon$ in the
	marginal massive case $\Delta^\infty_\F=-1$), and hence it vanishes as $\epsilon\to0$ whenever
	$\operatorname{Re}\Delta^\infty_\F<0$. Under this condition, the two straight segments combine into (minus) the
	discontinuity across the cut, yielding a representation of the Wick-rotated mode function as an integral over the cut alone:
	\begin{equation}\label{eq: dispersion Wick rotated}
		i\,\F(-iz) = -\int_{-\infty}^{-1}\frac{\d\lambda}{2\pi i}\,e^{\lambda z}\,
		\Disc_\lambda\big[\hat{\F}(\lambda)\big]\,.
	\end{equation}
	Changing the integration variable $\lambda\to-\lambda$ and rotating back $z\to iz$, we obtain
	\begin{equation}\label{eq: mode function dispersive 1}
		\boxed{\;
			\F(z) = \frac{1}{2\pi}\int_{1}^{\infty}\d\lambda\;
			e^{-i\lambda z}\,
			\big[\Disc_{\lambda'}\hat{\F}(\lambda')\big]_{\lambda'=-\lambda}\,.}
	\end{equation}
	This representation is valid for any bulk function $\F$ (with $\operatorname{Re}\Delta^\infty_\F<0$) and for $z$ lying in the intersection of the domain of analyticity of $\F$ and the lower half plane $\Im(z)<0$. The last condition ensures the convergence of the integral in \eqref{eq: mode function dispersive 1} and is consistent with the $i \epsilon$ prescription we started with in \eqref{eq: Fhat physical def}.  
	By \eqref{eq: disc reflected}, the $\lambda$-plane discontinuity can be traded for that of the mode
	function in the time domain; in what follows, however, we use \eqref{eq: mode function dispersive 1} directly, with the
	discontinuity supplied by the closed form \eqref{eq: Laplace transform massive closed} for the massive mode function.
	
	\paragraph{Plane-wave representation.}
	For the massive mode function the discontinuity is self-reproducing: the connection formulae
	for Legendre functions give, for $\lambda<-1$,
	\begin{equation}\label{eq: disc Legendre self}
		\Disc_\lambda\big[P_{i\mu-1/2}(\lambda)\big] = -2i\cosh(\pi\mu)\,P_{i\mu-1/2}(-\lambda)\,,
	\end{equation}
	so that $\big[\Disc_{\lambda'}\hat{\F}(\lambda')\big]_{\lambda'=-\lambda}=-2i\cosh(\pi\mu)\,
	\hat{\F}(\lambda)$. Inserting this together with
	\eqref{eq: Laplace transform massive closed} into
	\eqref{eq: mode function dispersive 1}, the $\cosh(\pi\mu)$ factors cancel and we obtain the
	plane-wave representation of the massive mode function,
	\begin{equation}\label{eq: massive mode plane wave}
		\boxed{\;\frac{H^{(1)}_{i\mu}(-z)}{\sqrt{-z}}
			= -\sqrt{\frac{2}{\pi}}\,e^{\frac{\pi\mu}{2}+\frac{i\pi}{4}}
			\int_1^\infty \d\lambda\;e^{-i\lambda z}\,P_{i\mu-1/2}(\lambda)\;}
	\end{equation}
	valid in the lower-half $z$ plane selected by the Bunch-Davies prescription (it also holds for $z \in \mathbb{R}^{-}$), and for any mass. The entire time
	dependence is now carried by the plane wave $e^{-i\lambda z}$, dressed by the Legendre weight
	$P_{i\mu-1/2}(\lambda)$ that encodes the mass. This is the central object on which the
	Laplace-space representation of correlators is built in Section~\ref{sec: correlators}.
	
	\paragraph{Consistency checks.}
	The representation \eqref{eq: massive mode plane wave} reduces to elementary mode
	functions in the two limits where the Legendre weight trivialises. For a conformally coupled
	field, $i\mu=\tfrac12$ and $P_0(\lambda)=1$; the prefactor reduces to $-\sqrt{2/\pi}$ and
	\begin{equation}
		\frac{H^{(1)}_{1/2}(-z)}{\sqrt{-z}}
		= -\sqrt{\frac{2}{\pi}}\int_1^\infty\d\lambda\;e^{-i\lambda z}
		= i\sqrt{\frac{2}{\pi}}\,\frac{e^{-iz}}{z}\,.
	\end{equation}
	For a massless field, $i\mu=\tfrac32$ and $P_1(\lambda)=\lambda$; the prefactor reduces to
	$i\sqrt{2/\pi}$ and carrying out the $\lambda$ integral gives
	\begin{equation}
		\frac{H^{(1)}_{3/2}(-z)}{\sqrt{-z}}
		= i\sqrt{\frac{2}{\pi}}\int_1^\infty\d\lambda\;\lambda\,e^{-i\lambda z}
		= \sqrt{\frac{2}{\pi}}\,\frac{z-i}{z^2}\,e^{-iz}\,,
	\end{equation}
	both reproducing the direct evaluation of the corresponding Hankel functions. Note that these cases lie precisely beyond the regime where the Laplace transform is initially defined, with $|\Im(\mu)| < 1/2$, and hence illustrate the reasoning described above based on analytic continuations.

	\paragraph{A general pattern.} Although derived here for the massive scalar in de Sitter space, the logic is generic: it applies to any mode
	function with Bunch-Davies asymptotics, and equations of motion with polynomial coefficients in $z$ readily give their Laplace-space counterparts. Several features remain
	unchanged. The Bunch-Davies behaviour implies regularity at the excited-state point $\lambda=1$ and a  branch point at $\lambda=1$, whose cut on $(-\infty,-1)$ is the dual image of the temporal discontinuity through
	\eqref{eq: disc reflected}; and the time evolution maps to a second-order equation in $\lambda$
	whose singular point $\lambda=-1$ carries the early-time data. What is \emph{not}
	universal is the local nature of the singularity at $\lambda=-1$, which reflects the field
	content. For a spin-1 field with a helical chemical potential the mode function is a Whittaker function
	$W_{i\tk,i\mu}$ rather than a Hankel function; the chemical potential term in its
	equation of motion promotes the dual Legendre equation to the \emph{associated} Legendre
	equation of order $i\tk$, whose solutions $P^{i\tk}_{i\mu-1/2}$ carry an extra factor
	$\big(\tfrac{\lambda+1}{\lambda-1}\big)^{i\tk/2}$ endowing $\lambda=-1$ with a non-integer
	branch exponent. The cut does not move, but the marginal logarithm of the de Sitter case is
	deformed. This de
	Sitter-breaking case and others are worked out in Appendix~\ref{app:Whittaker}.

	\subsection{Early-time resummation}\label{subsec_Borel}
	
	The plane-wave representation \eqref{eq: massive mode plane wave} also follows from a complementary and
	more physical argument, which we sketch before turning it into a proof. At early times, when a mode is deep inside the
	Hubble radius, the curvature is negligible and the mode function degenerates to its flat-space
	counterpart, which is an ordinary plane wave $e^{-ik\tau}$. The de Sitter
	geometry corrects this exact flat-space behaviour only through terms suppressed by powers of
	$1/(k\tau)$, organised into an early-time asymptotic expansion. That expansion is divergent but
	Borel summable: resumming it reconstructs the exact mode function as a continuous superposition
	of plane waves---the very representation \eqref{eq: massive mode plane wave} found above from the Laplace
	transform. We now make this precise.
	
	\paragraph{Borel resummation in a nutshell.}
	Borel resummation assigns a finite value to an asymptotic series whose coefficients grow
	factorially \cite{Dorigoni_2019,Aniceto_2019}.
    Given
	\begin{equation}
		f(g) \sim \sum_{n=0}^\infty c_n g^n\,,\qquad c_n \underset{n\to+\infty}{\sim} n!\,,
		\label{asymptotic_exp_generic}
	\end{equation}
	one divides out the factorial growth to form the Borel transform, which has a finite radius of
	convergence,
	\begin{equation}
		\mathcal{B}[f](\lambda) = \sum_{n=0}^\infty \frac{c_n}{n!}\,\lambda^n\,,
		\label{Borel_transform_generic}
	\end{equation}
	and recovers $f$ as its Borel sum, the Laplace transform of \eqref{Borel_transform_generic},
	\begin{equation}
		\mathcal{S} f(g) = \frac{1}{g} \int_0^\infty \d\lambda\, \mathcal{B}[f](\lambda)\,e^{-\lambda/g}\,.
		\label{Borel_sum_generic}
	\end{equation}
	The factor $e^{-\lambda/g}$ restores the non-perturbative dependence on $g$ that was lost in \eqref{asymptotic_exp_generic}, and the Borel sum
	analytically continues the series. If $\mathcal{B}[f]$ has poles on the positive
	axis, additional non-perturbative contributions arise with sign ambiguities that have to be fixed by physical or mathematical arguments, but no such ambiguity arises in the case at hand.

	\paragraph{Application to the massive function.}
	We now aim to use the Borel resummation process described above to find an integral representation of the massive mode function. The object of interest is the same as \eqref{eq:F-Hankel} in Section~\ref{subsec: dual equation}, here
	written in the cosmological normalisation for a field of momentum $s$:\footnote{We call $s$ the
		magnitude of the massive-field momentum to anticipate its role as an internal-line
		momentum in Section~\ref{sec: correlators}.}
	\begin{equation}
		\sigma(\tau,s)=-i\frac{\sqrt{\pi}}{2}H\,e^{-\frac{\pi}{2}\mu+i\frac{\pi}{4}}\,(-\tau)^{3/2}H_{i\mu}(-s\tau)\,,
		\label{eq:mode-function-massive}
	\end{equation}
	which obeys the equation of motion
	\begin{equation}
		\left(\partial_\tau^2 -\frac{2}{\tau}\partial_\tau+s^2+\frac{\mu^2+\frac{1}{4}}{\tau^2}\right)\sigma(\tau,s)=0\,.
		\label{eom_for_Borel}
	\end{equation}
	
	In the following, we compute the perturbative expansion of \eqref{eq:mode-function-massive} with respect to the dimensionless quantity $1/(-s\tau)$, i.e.\ the leading behaviour and its corrections in powers of $1/(-s\tau)$ up to all orders. From the asymptotics of the Hankel function, the leading-order early-time behaviour is given by $\sigma(\tau,s)\underset{\tau \to -\infty}{\sim} iH\tau \frac{e^{-i s\tau}}{\sqrt{2 s}}$.
	To compute the corrections, we consider the perturbative series
	\begin{equation}
		\sigma(\tau,s) \sim iH\tau\,\frac{e^{-i s\tau}}{\sqrt{2 s}} \sum_{n=0}^{\infty} \frac{\alpha_n(\mu)}{(-s\tau)^n}\;,\;\text{with}\;\alpha_0(\mu)=1 \,.
		\label{sigma_perturbative_series}
	\end{equation}
	Using this series as an ansatz in the differential equation \eqref{eom_for_Borel} verified by the mode function $\sigma$, we can find a recursion relation for the coefficients $\alpha_n(\mu)$:
	\begin{equation}
		\begin{aligned}
			\alpha_n(\mu) &= \left(-\frac{i}{2}\right) \frac{n(n-1)+\mu^2+\frac{1}{4}}{n}\, \alpha_{n-1}(\mu)\\
			&= \left(-\frac{i}{2}\right) \frac{\left(\frac{1}{2}+i\mu+n-1\right)\left(\frac{1}{2}-i\mu+n-1\right)}{n}\, \alpha_{n-1}(\mu)\,,
		\end{aligned}
	\end{equation}
	valid for $n\geq 1$, where the Pochhammer symbol is defined as $(a)_n \equiv\frac{\Gamma(a+n)}{\Gamma(a)}$. The recursion can be straightforwardly solved:
	\begin{equation}
		\alpha_n(\mu) = \left(-\frac{i}{2}\right)^n \frac{\Gamma(\frac{1}{2}+i\mu+n)\Gamma(\frac{1}{2}-i\mu+n)}{\Gamma(\frac{1}{2}-i\mu)\Gamma(\frac{1}{2}+i\mu)\Gamma(n+1)}=  \left(-\frac{i}{2}\right)^n \frac{\left(\frac{1}{2}-i\mu\right)_n \left(\frac{1}{2}+i\mu\right)_n}{n!}\,.
	\end{equation}
	Because $(a)_n \underset{n\rightarrow+\infty}{\sim} n!$, the coefficients diverge factorially, $\alpha_n(\mu)\underset{n\rightarrow+\infty}{\sim} n!$,
	hence the series \eqref{sigma_perturbative_series} is asymptotic. To summarise: the early-time perturbative expansion up to all orders reads $\sigma(\tau,s) \sim \sum_{n=0}^{\infty} c_n g^n$, with $g=-1/(s\tau)$ and
	\begin{equation}
		c_n = iH\tau\,\frac{e^{-i s\tau}}{\sqrt{2 s}}
		\left(-\frac{i}{2}\right)^n \frac{\left(\frac{1}{2}-i\mu\right)_n \left(\frac{1}{2}+i\mu\right)_n}{n!}\,.   
	\end{equation}
	From the knowledge of this asymptotic expansion, we can use the Borel resummation process, and use \eqref{Borel_sum_generic} to find
	\begin{equation}
		\sigma(\tau,s) = -iH\tau^2 \sqrt{\frac{s}{2}}\,e^{-i s\tau}
		\int_0^\infty \d \lambda\,e^{\lambda s \tau } \,_2F_1\left(\frac{1}{2}+i\mu,\frac{1}{2}-i\mu;1;-\frac{i\,\lambda}{2}\right)
		\,,
	\end{equation}
	where ${}_2F_1$ is the Gauss hypergeometric function (see Eq.~\eqref{eq: 2F1 def}).
	For $\tau$ in the lower-half plane, deforming the contour of integration to the negative imaginary axis, and shifting the integration variable, through $\lambda\to1+i\lambda$, so that
	$-\tfrac{i\lambda}{2}\to\tfrac{1-\lambda}{2}$ and $e^{\lambda s\tau}\to e^{is\tau}e^{-is\lambda\tau}$, we obtain
	\begin{equation}
		\sigma(\tau,s) = -H\sqrt{\frac{s}{2}}\tau^2 \int_1^\infty \d \lambda\,e^{-i s\lambda\tau} P_{i\mu-1/2}(\lambda)\,,
		\label{Borel_rep_massive}
	\end{equation}
	where the time and kinematic dependence of $\sigma/\tau^2$ is given by plane waves. This is fully equivalent to the representation~\eqref{eq: massive mode plane wave}, simply written in cosmological notation.

	A comment on the reach of the Borel construction is in order. It reconstructs a bulk mode function as a genuine superposition of plane waves $\int_1^\infty\d\lambda\,\rho(\lambda)\,e^{-i\lambda z}$ only when the early-time behaviour is of the form $e^{-iz}/z$: the resummation then packages the entire fall-off into the weight $\rho$. This is precisely what happens for $H^{(1)}_{i\mu}(-z)/\sqrt{-z}$, or equivalently $\sigma(\tau,s)/\tau^2$, with weight the Legendre function. When the early-time behaviour instead has an additional power-law dependence, $z^\beta e^{-iz}/z$---for instance as generated by a helical chemical potential through a non-integer twist exponent---that factor is not reorganised but pulled out of the integral, and the Borel sum returns $z^\beta$ times a superposition of plane waves. We carry this out explicitly for the general case of the twisted Whittaker mode function in Appendix~\ref{sec:Borel-Whittaker}, where the Borel representation is compared with the Laplace one.

	\section{From Flat Space to de Sitter Massive Cosmological Correlators}\label{sec: correlators}

	\subsection{Cosmological correlators}
	
	\paragraph{In-in correlators.} We are interested in equal-time correlators of field
	operators evaluated at a finite late time $\tau_0$,
	\begin{equation}
		\braket{\hat{O}(\tau_0)}\equiv\braket{\Omega|\hat{O}(\tau_0)|\Omega}\,,\qquad
		\hat{O}(\tau_0)=\prod_i \hat{\varphi}^{A_i}(\tau_0,\x_i)\,,
	\end{equation}
    where $\ket{\Omega}$ is the Bunch-Davies vacuum state. They are naturally expressed in the in-in (Schwinger-Keldysh) formalism~\cite{Weinberg_2005,Chen:2017ryl}. Expressing the vacuum-to-vacuum amplitude as a path integral
	requires two copies $\varphi_\pm$ of each field, propagating respectively forward and backward
	in time and glued at $\tau_0$. Adding external currents $J_\pm$, the generating functional reads
	\begin{equation}
		\begin{aligned}
			Z[J_+,J_-]=\int\mathcal{D}\varphi_+\mathcal{D}\varphi_-\,
			\exp\Bigg(&\,i\!\int_{-\infty^+}^{\tau_0}\!\!\d\tau\,\d^3\x\left(\mathcal{L}[\varphi_+]+J_+\varphi_+\right)\\
			&-i\!\int_{-\infty^-}^{\tau_0}\!\!\d\tau\,\d^3\x\left(\mathcal{L}[\varphi_-]+J_-\varphi_-\right)\Bigg)\,,
			\label{generating_functional_def}
		\end{aligned}
	\end{equation}
	with a delta function $\delta(\varphi_+(\tau_0)-\varphi_-(\tau_0))$ understood at the boundary.
	Correlators follow by functional differentiation,
	\begin{equation}
		\braket{\hat{\varphi}_{a_1}(\tau_0,\x_1)\ldots\hat{\varphi}_{a_N}(\tau_0,\x_N)}
		=\prod_{n=1}^N\frac{\delta}{i\,a_n\,\delta J_{a_n}(\tau_0,\x_n)}\,Z[J_+,J_-]\Big|_{J_\pm=0}\,,
		\label{correlators_from_Z}
	\end{equation}
	where each index $a_n=\pm$ labels the contour branch. In perturbation theory we split
	$\mathcal{L}=\mathcal{L}_2+\mathcal{L}_{\text{int}}$ and expand in $\mathcal{L}_{\text{int}}$. Taking the far-past limit projects onto the interacting vacuum and regularises
	the otherwise oscillating time integrals through a small tilt of the integration contour, $-\infty^\pm\equiv-\infty(1\mp i\epsilon)$, chosen on each branch so that the early-time oscillations are damped and the interacting vacuum
	is projected out. 
	
	\paragraph{Mode functions.} We consider a conformally coupled field $\varphi$, carrying the
	external legs, and an arbitrary set of massive fields $\sigma^A$ exchanged internally. The conformally
	coupled mode function is an exact plane wave up to a power factor,
	\begin{equation}
		\varphi(\tau;E)=\frac{iH\tau}{\sqrt{2E}}\,e^{-iE\tau}\,,
		\label{cc_mode_function}
	\end{equation}
	while the massive mode function is given in terms of a Hankel function:
	\begin{equation}\label{massive_mode_function_2}
		\sigma(\tau,s)=-i\frac{H\sqrt{\pi}}{2}e^{-\frac{\pi}{2}\mu+i\frac{\pi}{4}}(-\tau)^{3/2}H_{i\mu}^{(1)}(-s\tau)\,.
	\end{equation}
	The latter admits the
	plane-wave representation~\eqref{Borel_rep_massive} established in Section~\ref{sec: Laplace space}, valid for any real or imaginary value
	of $\mu=\sqrt{m^2/H^2-9/4}$, where $m$ is the mass. In the following, we denote generic mode functions by $u_k(\tau)$. 
	
	\paragraph{Propagators.} Differentiating the free generating functional twice yields the four
	Schwinger-Keldysh propagators. The $\pm\pm$ propagators are (anti-)time-ordered, while the
	$\pm\mp$ ones are Wightman functions,
	\begin{equation}
		\begin{aligned}
			&G_{++}(k,\tau_1,\tau_2)=\Theta(\tau_1-\tau_2)\,G_{>}+\Theta(\tau_2-\tau_1)\,G_{<}\,,
			&&G_{+-}(k,\tau_1,\tau_2)=G_{<}\,,\\
			&G_{--}(k,\tau_1,\tau_2)=\Theta(\tau_1-\tau_2)\,G_{<}+\Theta(\tau_2-\tau_1)\,G_{>}\,,
			&&G_{-+}(k,\tau_1,\tau_2)=G_{>}\,,
		\end{aligned}
		\label{def_propagators}
	\end{equation}
	with the Wightman functions built from the mode functions,
	\begin{equation}
		G_{>}(k,\tau_1,\tau_2)=u_k(\tau_1)\,u_k^*(\tau_2)\,,\qquad
		G_{<}(k,\tau_1,\tau_2)=u_k^*(\tau_1)\,u_k(\tau_2)\,.
		\label{def_G_<>}
	\end{equation}	
	Diagrammatically, an internal massive line carries a solid edge whose endpoints are decorated
	with a black dot for a $+$ vertex and a white dot for a $-$ vertex:
	\begin{equation}
		\begin{aligned}
			&G_{++} (s, \tau_1, \tau_2) = 
			\begin{tikzpicture}
				\filldraw (0,0) circle (2pt) node[above] {$\tau_1$};
				\filldraw (2,0) circle (2pt) node[above] {$\tau_2$};
				\draw[thick] (2,0) to (0,0);
				\node at (1,0.3) {$s$};
			\end{tikzpicture} \,, \\
			&G_{+-} (s, \tau_1, \tau_2) = 
			\begin{tikzpicture}
				\draw[thick] (2,0) to (0,0);
				\filldraw (0,0) circle (2pt) node[above] {$\tau_1$};
				\filldraw[color=black,fill=white] (2,0) circle (2pt) node[above] {$\tau_2$};
				\node at (1,0.3) {$s$};
			\end{tikzpicture} \,, \\
			&G_{-+} (s, \tau_1, \tau_2) = 
			\begin{tikzpicture}
				\draw[thick] (2,0) to (0,0);
				\filldraw[color=black,fill=white] (0,0) circle (2pt) node[above] {$\tau_1$};
				\filldraw (2,0) circle (2pt) node[above] {$\tau_2$};
				\node at (1,0.3) {$s$};
			\end{tikzpicture} \,, \\
			&G_{--} (s, \tau_1, \tau_2) =
			\begin{tikzpicture}
				\draw[thick] (2,0) to (0,0);
				\filldraw[color=black,fill=white] (0,0) circle (2pt) node[above] {$\tau_1$};
				\filldraw[color=black,fill=white] (2,0) circle (2pt) node[above] {$\tau_2$};
				\node at (1,0.3) {$s$};
			\end{tikzpicture} \,.
		\end{aligned}
		\label{diag_bulk_propagators}
	\end{equation}
	Bulk-to-boundary propagators have one endpoint pinned at $\tau_0$, drawn as a boundary square:
	\begin{equation}
		\begin{aligned}
			&K_+ (s, \tau) \equiv G_{++} (s, \tau, \tau_0) = 
			\begin{tikzpicture}
				\draw[thick] (2,0) to (0,0);
				\filldraw (0,0) circle (2pt) node[above] {$\tau$};
				\filldraw[color=black,fill=white,yshift=-2pt] (2,0) rectangle ++(4pt,4pt);
				\node at (1,0.3) {$s$};
			\end{tikzpicture} \,, \\
			&K_- (s, \tau) \equiv G_{-+} (s, \tau, \tau_0) =
			\begin{tikzpicture}
				\draw[thick] (2,0) to (0,0);
				\filldraw[color=black,fill=white] (0,0) circle (2pt) node[above] {$\tau$};
				\filldraw[color=black,fill=white,yshift=-2pt] (2,0) rectangle ++(4pt,4pt);
				\node at (1,0.3) {$s$};
			\end{tikzpicture} \,.
		\end{aligned}
		\label{diag_bdy_propagators}
	\end{equation}
	For the conformally coupled external legs, \eqref{cc_mode_function} gives the explicit
	bulk-to-boundary propagators
	\begin{equation}
		K_+(E,\tau)=\frac{H^2\tau_0\,\tau}{2E}\,e^{iE\tau}\,,\qquad
		K_-(E,\tau)=\frac{H^2\tau_0\,\tau}{2E}\,e^{-iE\tau}\,.
		\label{bulk_to_boundary_cc}
	\end{equation}

	\paragraph{Vertices.} Each occurrence of an interaction term in the perturbative expansion of
	\eqref{generating_functional_def} produces a vertex of either branch: a $+$ ($-$) vertex,
	drawn as a black (white) dot, contributes $+i\,S_{\text{int}}$ ($-i\,S_{\text{int}}$), i.e.\ a
	coupling, a conformal-time integral with the Bunch-Davies $i\epsilon$ prescription, and the
	appropriate scale-factor measure. For the two-field interaction
	$\mathcal{L}_{\text{int}}/a^4=-\frac{g}{n!\,p!}\,\varphi^n\sigma^p$, a vertex of branch
	$a_i=\pm$ evaluates to
	\begin{equation}
		\begin{tikzpicture}[baseline={(0,0)}]
			\draw[thick, black] (-1,1) to (0,0);
			\draw[thick, black] (-1,-1) to (0,0);
			\draw[thick, black] (-1.1,.6) to (0,0);
			\draw[thick, black] (0,0) to (1,1);
			\draw[thick, black] (1.1,.6) to (0,0);
			\draw[thick, black] (0,0) to (1,-1);
			\begin{scope}
				\clip (0,0) circle(2pt);
				\fill[black] (0,-2pt) rectangle (-2pt,2pt);
				\fill[white] (0,-2pt) rectangle (2pt,2pt);
			\end{scope}
			\draw[color=black] (0,0) circle (2pt)
			node[below] {$\tau$} node[above] {$a_i$};
			\node[black] at (-0.42,0.65) {$\varphi$};
			\node[black] at (0.42,0.65) {$\sigma$};
			\node[black] at (-.8,0) {$\ldots$};
			\node[black] at (.8,0) {$\ldots$};
		\end{tikzpicture}
		= -\,a_i\,i\,g\int_{-\infty_{a_i}}^{\tau_0}\!\d\tau\,a^4(\tau)
		\prod_{e=1}^{n}K_{a_i}(k_e;\tau)\prod_{j=1}^{p}G_{a_i a_j}(s_j;\tau,\tau_j)\,,
	\end{equation}
	with $a_j$ the branches of the neighbouring vertices.
	
	\paragraph{Assembling a diagram.} A correlator at a given order is the sum over the $2^V$ ways
	of assigning $\pm$ to the $V$ vertices; the all-reversed labelling of any contribution is its
	complex conjugate. Each labelled diagram is a product of $V$ nested time
	integrals over bulk-to-bulk and bulk-to-boundary propagators. The nesting is the central
	difficulty: any $\pm\pm$ internal line carries a time ordering, turning the corresponding
	integrals into layered ones. The strategy of the next subsection is to remove this obstruction
	by inserting the Laplace representation \eqref{Borel_rep_massive} of the massive mode function,
	which trades each massive line for a plane wave and reduces every time integral to a flat-space
	one.

	\subsection{From time representation} \label{diag_rules_correlators}
	
	\paragraph{Setup.}
	For definiteness, we consider the effective field theory of inflationary fluctuations in the decoupling limit and in a de Sitter background, with the Goldstone boson of spontaneously broken time translations $\pi$ coupled to an arbitrary number of scalar fields $\sigma^A$. The observables of interest are the correlation functions of $\pi$, which can be deduced from the correlators of a conformally coupled field $\varphi$ coupled to the $\sigma^A$, by acting with suitable weight-shifting operators and possibly taking soft limits of external momenta, see e.g.~\cite{Jazayeri:2022kjy}. For concreteness, we consider polynomial interactions and assume that all fields have a unit speed of sound, see Section~\ref{sec:generalisations} for generalisations.

	The two mode functions we need were recalled in the previous subsection: the conformally
	coupled one~\eqref{cc_mode_function} carried by external legs, with bulk-to-boundary propagators $K$~\eqref{bulk_to_boundary_cc}, and the massive
	one~\eqref{massive_mode_function_2}, carried by internal lines, together with its plane-wave
	representation~\eqref{Borel_rep_massive}.

	\paragraph{Inserting the Laplace representation.}
	Using the integral representation \eqref{Borel_rep_massive}, at a given vertex labelled with index $i$ the time dependence is given by plane waves up to an integer power factor. To determine this power factor in terms of the vertex features, consider the generic expression of quantities depending on the time coordinate associated with a vertex $i$, denoted $\tau_i$. We denote by $C_i$ the number of external conformally coupled lines attached to vertex $i$, and by $M_i$ the number of massive internal lines between vertex $i$ and any other vertex of the diagram. Therefore, all $\tau_i$-dependent quantities of the diagram are
	\begin{equation}
		\begin{aligned}
			&\begin{tikzpicture}[baseline={(0,0)}]
				\draw[thick, black] (-1.5,1.5) to (0,0);
				\draw[thick, black] (1.5,1.5) to (0,0);
				\filldraw[color=black,fill=white,xshift=-2pt,yshift=-2pt] (-1.5,1.5) rectangle ++(4pt,4pt);
				\filldraw[color=black,fill=white,xshift=-2pt,yshift=-2pt] (1.5,1.5) rectangle ++(4pt,4pt);
				\node at (0,1.2) {$\ldots$};
				\draw[thick, black] (-1.5,-1.5) to (0,0);
				\draw[thick, black] (1.5,-1.5) to (0,0);
				\node[black] at (-1.1,0.6) {$\k_1$};
				\node[black] at (1.1,0.6) {$\k_{C_i}$};
				\node[black] at (-1.1,-.6) {$s_{i\,1}$};
				\node[black] at (1.1,-.6) {$s_{i\,M_i}$};
				\node[black] at (0,-1.2) {$\ldots$};
				\begin{scope}
					\clip (0,0) circle(2pt);
					\fill[black] (0,-2pt) rectangle (-2pt,2pt);
					\fill[white] (0,-2pt) rectangle (2pt,2pt);
				\end{scope}
				\draw[color=black] (0,0) circle (2pt) node[above] {$\tau_i$};
			\end{tikzpicture}
			&= - i a_i g_i \int_{-\infty_{a_i}}^0 \d\tau_i\, a^4(\tau_i) \prod_{e=1}^{C_i} K_{a_i}(k_e;\tau_i)\;\prod_{j=1}^{M_i} G_{a_i\,a_j}(s_{ij};\tau_i,\tau_j)\,,
		\end{aligned}
		\label{eq:initial-generic-vertex}    
	\end{equation}
	where $g_i$ is the coupling constant of the vertex $i$, $a_i$ is its Schwinger-Keldysh index and the $a_j$ are those of neighbouring ones. The first product is over external legs, each carrying momentum $\k_e$ with norm $k_e$. For $1 \leq j \leq M_i$, each internal line accounts for the propagation of a field of mass $\mu_{ij}$ and momentum norm $s_{ij}$. Now, in propagators $G$ on the right-hand side of the last expression, we write the $M_i$ massive mode functions attached to vertex $i$ using \eqref{Borel_rep_massive} and associate to each one an integration parameter $\lambda_{ij}$. In the notation $\lambda_{ij}$, the first subscript corresponds to the attached vertex and the second one is the label of the other vertex connected by the internal line. Therefore, the above expression becomes: 
	\begin{equation}
		\begin{aligned}
			- &\frac{i a_i g_i\, H^{2 C_i -4 + 2 M_i} \,\tau_0^{C_i}}{\left(\prod_{e=1}^{C_i} \,2 k_e\right) 2^{M_i/2}}
			\int_{-\infty}^0 \d\tau_i\,\tau_i^{C_i-4+2 M_i}\,e^{i a_i E_i\tau_i}\\
			&\times\prod_{j=1}^{M_i} s_{ij}^{1/2} \int_1^\infty \d\lambda_{ij} \,P_{i\mu_{ij}-\frac{1}{2}}(\lambda_{ij})\;G_{a_i\,a_j}^{\textrm{flat}}(s_{ij};\tau_i,\tau_j;\lambda_{ij},\lambda_{ji})\,,
		\end{aligned}
		\label{vertex_i_lambda}
	\end{equation}
	with the time dependence of the scale factor and of bulk-to-boundary propagators being written explicitly, and where
	\begin{equation}
		E_i\equiv \sum_{e=1}^{C_i} k_e\,.
	\end{equation}
	In the expression \eqref{vertex_i_lambda}, for the $M_i$ massive mode functions included in propagators $G_{a_i\,a_j}$ but attached to neighbouring vertices, we included their plane-wave dependence resulting from the use of \eqref{Borel_rep_massive} in functions $G_{a_i\,a_j}^{\textrm{flat}}(s_{ij};\tau_i,\tau_j;\lambda_{ij},\lambda_{ji})$.
	These functions are similar to flat-space propagators with a rescaling of the exchanged momentum $s_{ij}$ by the Laplace parameters, either $\lambda_{ij}$ or $\lambda_{ji}$ depending on the mode function:
	\begin{eqnarray}
		G_{++}^{\textrm{flat}}(s_{ij};\tau_i,\tau_j;\lambda_{ij},\lambda_{ji})
		&=& \Theta(\tau_j-\tau_i) \, e^{i s_{ij} \lambda_{ij}(1-i \epsilon) \tau_i} e^{-i s_{ij} \lambda_{ji}(1+i\epsilon) \tau_j}+(i \leftrightarrow j) \nonumber \\
		&=&\left(G_{--}^{\textrm{flat}}(s_{ij};\tau_i,\tau_j;\lambda_{ij},\lambda_{ji})\right)^* \label{flat_G++}
		\\
		G_{+-}^{\textrm{flat}}(s_{ij};\tau_i,\tau_j;\lambda_{ij},\lambda_{ji})
		&=& e^{i s_{ij} \lambda_{ij}(1-i \epsilon) \tau_i} e^{-i s_{ij} \lambda_{ji}(1+i \epsilon) \tau_j} \nonumber \\
		&=&\left(G_{-+}^{\textrm{flat}}(s_{ij};\tau_i,\tau_j;\lambda_{ij},\lambda_{ji})\right)^*
		\,.
		\label{flat_G+-}
	\end{eqnarray}
	Let us highlight that the contour prescription in the $\tau_i$ integral has disappeared when going from \eqref{eq:initial-generic-vertex} to \eqref{vertex_i_lambda}. Instead, we trade it for an explicit deformation of the $\lambda$ variables in $G_{a_i\,a_j}^{\textrm{flat}}$, which are multiplied by $(1\pm i \epsilon)$.
	These regulators ensure that each time integral has an additional $e^{\epsilon \tau}$ convergent factor. After time integration, this will manifest itself as complex deformations of the dual variables $\lambda_{ij}$, the Laplace-space avatar of the Schwinger-Keldysh contour prescriptions.
	
	For further convenience, we define, from a diagram $\mathcal{F}\left(\{k_e\},\{s_{ij}\}\right)$, its rescaled counterparts $\tilde{\mathcal{F}}_{\{a_i\}}\left(\{E_i\},\{s_{ij}\}\right)$ by extracting the following factor: 
	\begin{equation}
		\mathcal{F}\left(\{k_e\},\{s_{ij}\}\right)\equiv \sum_{a_i=\pm}
		\left[\prod_{i\in\mathcal{V}} 
		-\frac{i a_i g_i\, H^{2 C_i -4 + 2 M_i} \,\tau_0^{C_i}}{\left(\prod_{e=1}^{C_i} \,2 k_e\right) 2^{M_i/2}} \prod_{i,j\in\mathcal{V};i<j} s_{ij}\right]
		\tilde{\mathcal{F}}_{\{a_i\}}\left(\{E_i\},\{s_{ij}\}\right)\,,
		\label{def_rescaled_diag}
	\end{equation}
	where $\mathcal{V}$ is the set of vertices. We will only refer to rescaled diagrams $\tilde{\mathcal{F}}_{\{a_i\}}$ in the following, corresponding to a given SK contribution with fixed $a_i=\pm$.
	
	\paragraph{Plane waves only.} In the integral over $\tau_i$ in \eqref{vertex_i_lambda}, in addition to the plane waves in $G_{a_i\,a_j}^{\textrm{flat}}$, there is also a power-law component $\tau_i^{N_i}$, with the integer 
	\begin{equation}
		N_i=C_i-4+2 M_i\,.
		\label{power_exponent_vertex}
	\end{equation}
	Our aim here is to show that, in practice, one can always bring the computation to a situation where $N_i=0$. For this, let us examine in turn the two other possibilities, $N_i>0$ or $N_i <0$.

	\begin{itemize}
		\item \textbf{If $N_i>0$}, observe that
		$\tau_i^{N_i}e^{\pm i E_i \tau_i}=\left(\mp i\right)^{N_i}\frac{\partial}{\partial E_i^{N_i}}\left(e^{\pm i E_i \tau_i}\right)$.
		Therefore, for a given SK contribution, we can simply compute a seed integral with $\tau_i^{N_i}$ removed, and recover $\tilde{\mathcal{F}}_{\{a_i\}}$ by applying the differential operator  $\left(- a_i \,i\right)^{N_i}\frac{\partial}{\partial E_i^{N_i}}$,
		where $a_i=\pm$ is the SK index of the vertex $i$. Note that it was important here to consider rescaled quantities, since the dependence of $\tilde{\mathcal{F}}_{a_i}$ on external kinematics is only through the plane wave $e^{i a_i E_i \tau_i}$ coming from the conformally coupled bulk-to-boundary propagators \eqref{bulk_to_boundary_cc}.
		
		Furthermore, we must also pay attention to the case where $E_i=0$ (e.g.\ a vertex without conformally coupled external lines, connected only to several massive internal lines). In that case, we artificially introduce a regulator $e^{i E_i \tau_i}$ in the seed time integral, for small $E_i>0$, and exchange the limit $E_i\rightarrow 0$ and integration over $\tau_i$. Note that the limit and the integration can be safely exchanged, thanks to the $i\epsilon$-prescription that ensures the convergence of the integral. The process described above can then be applied.   
		In summary, we apply $\lim_{E_i\rightarrow 0}\,\left(- a_i \,i\right)^{N_i}\frac{\partial}{\partial E_i^{N_i}}$ to the seed integral obtained by dropping the $\tau_i^{N_i}$ factor and adding the regulator $e^{i E_i \tau_i}$ to the rescaled diagram.
		\item \textbf{If $N_i<0$}, it is no longer possible to differentiate a seed integral and immediately obtain the desired diagram. One has $N_i\geq0$ if the vertex has two or more massive internal legs ($M_i\geq 2$). The only diagram with $N_i<0$ is for $M_i=1$ and $C_i=1$, giving $N_i=-1$ and corresponding to a quadratic mixing $\propto \varphi \sigma^A$. However, one can always bypass considering this situation. Indeed, in the EFT of inflationary fluctuations,
		the quadratic mixing vertex between the Goldstone boson $\pi$ and a massive degree of freedom $\sigma$ that is $\propto\pi'\sigma$ can be computed from an interaction vertex $\propto \sigma\,\varphi^2$ by taking a soft limit in one of the two conformally coupled external legs, see e.g.~\cite{Jazayeri:2022kjy}. This vertex has a number $N_i\geq 0$, to which we can apply the method described above.
	\end{itemize}
	Repeating this counting process for each vertex provides a relation between the diagram under consideration and seed time integrals over plane waves only. From this last quantity, the desired diagram $\tilde{\mathcal{F}}_{a_i}$ is obtained by taking derivatives (and possibly limits) at the end of the calculation.
	Moreover, each massive mode function attached to the vertex $i$ and to the internal line connecting vertices $i$ and $j$ brings an integral over a dual variable $\lambda_{ij}$ with integration measure $\int_1^\infty \d \lambda_{ij}\,P_{i\mu-1/2}(\lambda_{ij})$.

	\subsection{To Laplace representation}
	
	We now perform the remaining flat-space time integrals over plane waves in simple examples. This will make transparent the rules that allow us to write the Laplace-space integrand directly from a diagrammatic representation of correlators.

	\paragraph{Single exchange.} Let us first consider the $++$ component of the single-exchange diagram. Taking into account the two pieces of the time-ordered propagator \eqref{flat_G++}, we write its contribution coming from the time integrals as the following sum:
	\begin{equation}
		\begin{tikzpicture}[baseline={(0,0)}]
			\filldraw (0,0) circle (2pt) node[above] {$E_1$} node[anchor=north west] {$\lambda_{12}$};
			\filldraw (2,0) circle (2pt) node[above] {$E_2$} node[anchor=north east] {$\lambda_{21}$};;
			\draw[thick] (0,0) to (2,0);
			\node at (1,0.3) {$s_{12}$};
		\end{tikzpicture}=
		\begin{tikzpicture}[baseline={(0,0)}]
			\tikzset{thick curved arrow/.style={
					thick,
					decoration={markings, mark=at position 0.55 with {\arrow[scale=1.5]{latex'}}},
					postaction={decorate}
			}}
			\filldraw (0,0) circle (2pt) node[above] {$E_1$} node[anchor=north west] {$\lambda_{12}$};
			\filldraw (2,0) circle (2pt) node[above] {$E_2$} node[anchor=north east] {$\lambda_{21}$};;
			\draw[thick curved arrow] (0,0) to (2,0);
			\node at (1,0.3) {$s_{12}$};
		\end{tikzpicture}
		+\begin{tikzpicture}[baseline={(0,0)}]
			\tikzset{thick curved arrow/.style={
					thick,
					decoration={markings, mark=at position 0.55 with {\arrow[scale=1.5]{latex'}}},
					postaction={decorate}}}
			\filldraw (0,0) circle (2pt) node[above] {$E_1$} node[anchor=north west] {$\lambda_{12}$};
			\filldraw (2,0) circle (2pt) node[above] {$E_2$} node[anchor=north east] {$\lambda_{21}$};;
			\draw[thick curved arrow] (2,0) to (0,0);
			\node at (1,0.3) {$s_{12}$};
		\end{tikzpicture}
		\,,
	\end{equation}
	where each term is given by
	\begin{subequations}
		\begin{equation}
			\begin{aligned}
				\begin{tikzpicture}[baseline={(0,0)}]
					\tikzset{thick curved arrow/.style={
							thick,
							decoration={markings, mark=at position 0.55 with {\arrow[scale=1.5]{latex'}}},
							postaction={decorate}
					}}
					\filldraw (0,0) circle (2pt) node[above] {$E_1$} node[anchor=north west] {$\lambda_{12}$};
					\filldraw (2,0) circle (2pt) node[above] {$E_2$} node[anchor=north east] {$\lambda_{21}$};;
					\draw[thick curved arrow] (0,0) to (2,0);
					\node at (1,0.3) {$s_{12}$};
				\end{tikzpicture}
				& \equiv \int_{-\infty}^0 \d\tau_1 \int_{-\infty}^0 \d\tau_2 \, e^{i E_1 \tau_1} e^{i E_2 \tau_2} \,\Theta(\tau_2-\tau_1) \, e^{i s_{12} \lambda_{12} \tau_1+\epsilon \tau_1} e^{-i s_{12} \lambda_{21} \tau_2+\epsilon \tau_2}\,,
			\end{aligned}
			\label{integral_single_exch_1_2}
		\end{equation}
		\begin{equation}
			\begin{aligned}
				\begin{tikzpicture}[baseline={(0,0)}]
					\tikzset{thick curved arrow/.style={
							thick,
							decoration={markings, mark=at position 0.55 with {\arrow[scale=1.5]{latex'}}},
							postaction={decorate}}}
					\filldraw (0,0) circle (2pt) node[above] {$E_1$} node[anchor=north west] {$\lambda_{12}$};
					\filldraw (2,0) circle (2pt) node[above] {$E_2$} node[anchor=north east] {$\lambda_{21}$};;
					\draw[thick curved arrow] (2,0) to (0,0);
					\node at (1,0.3) {$s_{12}$};
				\end{tikzpicture}
				& \equiv  \int_{-\infty}^0 \d\tau_1 \int_{-\infty}^0 \d\tau_2 \, e^{i E_1 \tau_1} e^{i E_2 \tau_2} \,\Theta(\tau_1-\tau_2) \, e^{-i s_{12} \lambda_{12} \tau_1+\epsilon \tau_1} e^{i s_{12} \lambda_{21} \tau_2+\epsilon \tau_2}
				\,,
			\end{aligned}
			\label{integral_single_exch_2_1}
		\end{equation}
		\label{integral_single_exch_nested}
	\end{subequations}
	and, consistent with \eqref{vertex_i_lambda}, the vertex integral over $\tau_i$ includes the factor $e^{i\tau_i E_i}$ due to the presence of conformally coupled external legs, or of the regulator described in the previous section. Here and in the following, we do not show conformally coupled bulk-to-boundary lines, but rather indicate only the total external energy flowing at each vertex, here $E_1$ and $E_2$.

	Considering \eqref{integral_single_exch_1_2} for definiteness, integration over the earliest time $\tau_1$ gives
	\begin{equation}
		\int_{-\infty}^{\tau_2} \d\tau_1\,e^{i E_1 \tau_1}\, e^{i s_{12} \lambda_{12} \tau_1}e^{\epsilon \tau_1}=\frac{1}{iV_1+ \epsilon} e^{i V_1 \tau_2}\,,
	\end{equation}
	where
	\begin{equation}
		V_1= E_1+s_{12} \lambda_{12}\,
	\end{equation}
	involves the external energy $E_1$ and the rescaled internal energy $s_{12}\lambda_{12}$.
	Similarly performing the $\tau_2$ integral leads to
	\begin{equation}
		\begin{tikzpicture}[baseline={(0,0)}]
			\tikzset{thick curved arrow/.style={
					thick,
					decoration={markings, mark=at position 0.55 with {\arrow[scale=1.5]{latex'}}},
					postaction={decorate}
			}}
			\filldraw (0,0) circle (2pt) node[above] {$E_1$} node[anchor=north west] {$\lambda_{12}$};
			\filldraw (2,0) circle (2pt) node[above] {$E_2$} node[anchor=north east] {$\lambda_{21}$};;
			\draw[thick curved arrow] (0,0) to (2,0);
			\node at (1,0.3) {$s_{12}$};
		\end{tikzpicture}
		= \frac{1}{iV_1+ \epsilon}\int_{-\infty}^0 \d\tau_2 \, e^{i (V_1+V_2) \tau_2}e^{\epsilon \tau_2}=\frac{1}{i V_1+ \epsilon}\frac{1}{i(V_1+V_2)+ \epsilon}\,
		\label{integral_single_exch_1_2_res}
	\end{equation}
	where 
	\begin{equation}
		V_2= E_2-s_{12} \lambda_{21}\,.
	\end{equation}

	As for the flat-space integral corresponding to the $+-$ component of the single-exchange diagram, the unordered building block \eqref{flat_G+-} immediately gives
	\begin{equation}
		\begin{tikzpicture}[baseline={(0,0)}]
			\filldraw (0,0) circle (2pt) node[above] {$E_1$} node[anchor=north west] {$\lambda_{12}$};
			\draw (2,0) circle (2pt) node[above] {$E_2$} node[anchor=north east] {$\lambda_{21}$};;
			\draw[thick] (2,0) to (0,0);
			\node at (1,0.3) {$s_{12}$};
		\end{tikzpicture}
		=\int_{-\infty}^0 \d\tau_1 \int_{-\infty}^0 \d\tau_2 \, e^{i E_1 \tau_1} e^{-i E_2 \tau_2} \,e^{i s_{12} \lambda_{12}(1-i \epsilon) \tau_1} e^{-i s_{12} \lambda_{21}(1+i \epsilon) \tau_2}
		\,,
		\label{integral_single_exch_+-}
	\end{equation}
	leading to the factorised contribution:
	\begin{equation}
		\begin{tikzpicture}[baseline={(0,0)}]
			\filldraw (0,0) circle (2pt) node[above] {$E_1$} node[anchor=north west] {$\lambda_{12}$};
			\draw (2,0) circle (2pt) node[above] {$E_2$} node[anchor=north east] {$\lambda_{21}$};;
			\draw[thick] (2,0) to (0,0);
			\node at (1,0.3) {$s_{12}$};
		\end{tikzpicture}
		= \frac{1}{i(E_1+s_{12} \lambda_{12})+ \epsilon}\frac{1}{-i(E_2+s_{12} \lambda_{21})+ \epsilon}\,.
		\label{integral_single_exch_+-_res}
	\end{equation}

	This example of the single-exchange diagram illustrates the general lesson that, due to the time-ordering, the rational fraction obtained by integrating over the time coordinate of a given vertex includes those of past vertices, i.e.\ vertices connected by an ingoing arrow, and hence that the result can be simply obtained recursively by following time arrows.
	Moreover, for $+$-type (i.e.\ black-dotted) vertices, each line contributes either $+s\lambda$ for outgoing arrows and unordered lines or $-s\lambda$ for ingoing arrows, $s$ being the momentum norm carried and $\lambda$ the relevant Laplace parameter.
	
	\paragraph{Double exchange.}  Situations in which a vertex has at least two outgoing arrows come with a new ingredient. To see this, consider the following ordered contribution to the fully nested double-exchange diagram (omitting the $\epsilon$ terms to avoid clutter):
	\begin{equation}
		\begin{aligned}
			\begin{tikzpicture}[baseline={(0,0)}]
				\tikzset{thick curved arrow/.style={
						thick,
						decoration={markings, mark=at position 0.55 with {\arrow[scale=1.5]{latex'}}},
						postaction={decorate}}}
				\filldraw (0,0) circle (2pt) node[above] {$E_1$} node[anchor=north west] {$\lambda_{12}$};
				\filldraw (2,0) circle (2pt) node[above] {$E_2$} node[anchor=north east] {$\lambda_{21}$} node[anchor=north west] {$\lambda_{23}$};
				\filldraw (4,0) circle (2pt) node[above] {$E_3$} node[anchor=north east] {$\lambda_{32}$};
				\draw[thick curved arrow] (2,0) to (0,0);
				\draw[thick curved arrow] (2,0) to (4,0);
				\node at (1,0.3) {$s_{12}$};
				\node at (3,0.3) {$s_{23}$};
			\end{tikzpicture}
			&= \int_{-\infty}^0 \d\tau_1 \int_{-\infty}^0 \d\tau_2 \int_{-\infty}^0 \d\tau_3 \, e^{i E_1 \tau_1} e^{i E_2 \tau_2} e^{i E_3 \tau_3} \\
			&\hspace{- 1 cm}\times\,\Theta(\tau_1-\tau_2)  e^{-i s_{12} \lambda_{12} \tau_1}\,e^{i s_{12} \lambda_{21} \tau_2}\,\Theta(\tau_3-\tau_2) \, e^{i s_{23} \lambda_{23} \tau_2} e^{-i s_{23} \lambda_{32} \tau_3}\,.
		\end{aligned}
		\label{double_exch_int_T_2_1_3}
	\end{equation}
	To compute the integral over $\tau_2$, we have to specify the ordering of vertices $1$ and $3$. Hence, we write
	\begin{equation}
		1=\Theta(\tau_1-\tau_3)+\Theta(\tau_3-\tau_1)\,.
	\end{equation}
	With respect to the quantities $V_i$, defined here as
	\begin{equation}
		\begin{aligned}
			&V_1=E_1-s_{12} \lambda_{12}\\
			&V_2=E_2+s_{12} \lambda_{21}+s_{23} \lambda_{23}\\
			&V_3=E_3-s_{23} \lambda_{32}\,,
		\end{aligned}
		\label{V_123_2}
	\end{equation}
	this ordered diagram contribution is hence
	\begin{equation}
    \begin{aligned}
		\begin{tikzpicture}[baseline={(0,0)}]
			\tikzset{thick curved arrow/.style={
					thick,
					decoration={markings, mark=at position 0.55 with {\arrow[scale=1.5]{latex'}}},
					postaction={decorate}}}
			\filldraw (0,0) circle (2pt) node[above] {$E_1$} node[anchor=north west] {$\lambda_{12}$};
			\filldraw (2,0) circle (2pt) node[above] {$E_2$} node[anchor=north east] {$\lambda_{21}$} node[anchor=north west] {$\lambda_{23}$};
			\filldraw (4,0) circle (2pt) node[above] {$E_3$} node[anchor=north east] {$\lambda_{32}$};
			\draw[thick curved arrow] (2,0) to (0,0);
			\draw[thick curved arrow] (2,0) to (4,0);
			\node at (1,0.3) {$s_{12}$};
			\node at (3,0.3) {$s_{23}$};
		\end{tikzpicture}
		&= \frac{1}{i V_2+ \epsilon}\left(\frac{1}{i (V_1+V_2)+ \epsilon}+\frac{1}{i (V_1+V_3)+ \epsilon}\right) \nonumber \\
        & \times \frac{1}{i (V_1+V_2+V_3)+ \epsilon}\,.\label{time_int_T_2_1_3_res}
         \end{aligned}
	\end{equation}
	This example illustrates the fact that when we consider a vertex that has two or more outgoing arrows, we have to sum over all possible time orderings of later vertices.
	
	\paragraph{Loops.} Diagrams including loops can be considered in the same framework without modification, e.g.~the flat-space integral of the following loop diagram simply reads 
	\begin{equation}
		\begin{tikzpicture}[baseline={(0,0)}]
			\tikzset{thick curved arrow/.style={
					thick,
					decoration={markings, mark=at position 0.55 with {\arrow[scale=1.5]{latex'}}},
					postaction={decorate}}}
			\filldraw[color=black] (0,0) circle (2pt) node[left] {$E_1$};
			\filldraw[color=black] (60pt,0) circle (2pt) node[right] {$E_2$};
			\draw[thick curved arrow] (0,0) arc (180:0:30pt) node[midway,above] {$s_a$};
			\draw[thick curved arrow] (0,0) arc (-180:0:30pt) node[midway,below] {$s_b$};
			\node at (0,25pt) {$\lambda_{1a}$};
			\node at (0pt,-25pt) {$\lambda_{1b}$};
			\node at (60pt,25pt) {$\lambda_{2a}$};
			\node at (60pt,-25pt) {$\lambda_{2b}$};
		\end{tikzpicture}\,=
		\frac{1}{i\,V_1+\epsilon}\frac{1}{i\,\left(V_1+V_2\right)+\epsilon}\,,
	\end{equation}
	where
	\begin{equation}
		V_1 = E_1+s_a \lambda_{1a}+s_b\lambda_{1b}\,,\quad V_2= E_2-s_a \lambda_{2a}-s_b\lambda_{2b}\,.
	\end{equation}
	As usual, one simply has to add the integration over the corresponding loop momenta, and discard closed time-lines, which give vanishing contributions, see e.g.~\cite{AguiSalcedo:2023nds}.

	\subsection{Diagrammatic rules in Laplace space}
	\label{sec:rules-Laplace}
	
	We can now gather and summarise the complete set of Laplace-space diagrammatic rules that allow us to write directly any correlator in terms of $\lambda$-space integrals.
	
	\begin{enumerate}
		\item Decompose the diagram under consideration into SK contributions (i.e.\ assign $\pm$ to each vertex). Then, label each vertex with a number $i$ and compute the number
		\begin{equation}
			N_i=C_i-4+2 M_i\,,
			\label{power_exponent_vertex_summary}
		\end{equation}
		where $C_i$ is the number of external conformally coupled legs and $M_i$ is the number of internal massive lines connected to $i$. We always have $N_i\geq 0$ for physically relevant diagrams. Then, apply the differential operator
		\begin{equation}
			\left(- a_i \,i\right)^{N_i}\frac{\partial^{N_i}}{\partial E_i^{N_i}}\,,
			\label{diff_op_Ei_summary}
		\end{equation}
		at the end of the computation. If there is no external leg connected to the vertex $i$, introduce the vertex energy $E_i>0$ as a regulator (without modification of the number $N_i$), apply the previous differential operator and take the limit $E_i\rightarrow 0$.
		\item Assign the dual variable $\lambda_{ij}$ to the vertex $i$ and the internal line connecting vertices $i$ and $j$. Decompose each $\pm\pm$ internal line as
		\begin{equation}
			\begin{tikzpicture}[baseline={(0,0)}]
				\filldraw (0,0) circle (2pt) node[above] {$E_i$} node[anchor=north west] {$\lambda_{ij}$};
				\filldraw (2,0) circle (2pt) node[above] {$E_j$} node[anchor=north east] {$\lambda_{ji}$};
				\draw[thick] (0,0) to (2,0);
				\node at (1,0.3) {$s_{ij}$};
			\end{tikzpicture}=
			\begin{tikzpicture}[baseline={(0,0)}]
				\tikzset{thick curved arrow/.style={
						thick,
						decoration={markings, mark=at position 0.55 with {\arrow[scale=1.5]{latex'}}},
						postaction={decorate}
				}}
				\filldraw (0,0) circle (2pt) node[above] {$E_i$} node[anchor=north west] {$\lambda_{ij}$};
				\filldraw (2,0) circle (2pt) node[above] {$E_j$} node[anchor=north east] {$\lambda_{ji}$};
				\draw[thick curved arrow] (0,0) to (2,0);
				\node at (1,0.3) {$s_{ij}$};
			\end{tikzpicture}
			+
			\begin{tikzpicture}[baseline={(0,0)}]
				\tikzset{thick curved arrow/.style={
						thick,
						decoration={markings, mark=at position 0.55 with {\arrow[scale=1.5]{latex'}}},
						postaction={decorate}}}
				\filldraw (0,0) circle (2pt) node[above] {$E_i$} node[anchor=north west] {$\lambda_{ij}$};
				\filldraw (2,0) circle (2pt) node[above] {$E_j$} node[anchor=north east] {$\lambda_{ji}$};
				\draw[thick curved arrow] (2,0) to (0,0);
				\node at (1,0.3) {$s_{ij}$};
			\end{tikzpicture}
			\label{eq:decomposition-ordered-lambda}
		\end{equation}
		A SK diagram contribution that has $m$ $\pm\pm$ internal lines therefore decomposes into $2^m$ ordered contributions. We also define time-lines, which are sequences of internal lines obtained by following arrows. When considering a loop diagram, contributions involving closed time-lines vanish.
		\item For each variable $\lambda_{ij}$, write the following integral and integration measure (a diagram with $m$ internal lines is then expressed in terms of $2 m$ $\lambda$-integrals):
		\begin{equation}
			\int_1^\infty \d \lambda_{ij}\,P_{i\mu_{ij}-1/2}(\lambda_{ij})\,,
		\end{equation}
		where $\mu_{ij}$ is the mass parameter of the field exchanged between vertices $i$ and $j$.
		\item For each diagram contribution, compute the $\lambda$-space integrand:
		\begin{itemize}
			\item For a vertex $i$ with SK index $a_i$, compute the quantity $V_i$ that is the sum of the following contributions:
			\begin{itemize}
				\item add the vertex energy $+E_i$;
				\item for an outgoing arrow or an unordered line related to a vertex $j$, add a $+s_{ij}\,\lambda_{ij}$ factor;
				\item for an ingoing arrow related to a vertex $j$, add a $-s_{ij}\,\lambda_{ij}$ factor.
			\end{itemize}
			\item Compute $\mathcal{V}_i =i a_i (V_i +\sum_j V_j)+\epsilon$ where the sum is over earlier vertices (i.e.\ vertices related to $i$ by ingoing arrows in a time-line).
			\item If two or more time-lines are crossing without ending at a vertex, sum over all possible time orderings of later vertices. For a given ordering, add to the computation of $\mathcal{V}_i$ contributions $V_j$ from earlier vertices according to this ordering.
			\item The $\lambda$-space integrand is given by the product over vertices $\prod_i \frac{1}{\mathcal{V}_i}$.
		\end{itemize}
		\item The Laplace-space expression of a given SK contribution is then
		\begin{equation}
			\begin{aligned}
				\tilde{\mathcal{F}}_{\{a_i\}}\left(\{E_i\},\{s_{ij}\}\right)= &\int_1^{+\infty} \left(\prod_{i,j\in\mathcal{V},i\neq j} \d\lambda_{ij} \,P_{i \mu_{ij} -\frac{1}{2}}(\lambda_{ij} )\right)\\
				&\left(\prod_{i\in\mathcal{V}} \left(- a_i \,i\right)^{N_i}\frac{\partial}{\partial E_i^{N_i}}
				\;\sum_{\text{time ordering}} \frac{1}{\mathcal{V}_i(E_i,s_{ij},\lambda_{ij})}\right)\,,
			\end{aligned}
			\label{general_lambda_diagram}
		\end{equation}
		where $\mathcal{V}$ is the set of vertices and the sum is over all possibilities in the decomposition \eqref{eq:decomposition-ordered-lambda} for each $\pm\pm$ internal line. For a loop diagram, add loop momenta integrals, and modify the first parenthesis to incorporate dual variables for each massive mode propagating in each loop internal line. Finally, recover the initial diagram expression $\mathcal{F}_{\{a_i\}}\left(\{k_e\},\{s_{ij}\}\right)$ using \eqref{def_rescaled_diag}.
	\end{enumerate}

	\subsection{A master series for the single-exchange correlator}
	\label{subsec: single exchange}

	Let us assemble all pieces for the simple example of the following single-exchange correlator:
	\begin{equation}
		\begin{tikzpicture}[baseline={(0,0)}]
			\tikzset{thick curved arrow/.style={
					thick,
					decoration={markings, mark=at position 0.55 with {\arrow[scale=1.5]{latex'}}},
					postaction={decorate}}}
			\begin{scope}
				\clip (0,0) circle(2pt);
				\fill[black] (0,-2pt) rectangle (-2pt,2pt);
				\fill[white] (0,-2pt) rectangle (2pt,2pt);
			\end{scope}
			\draw (0,0) circle (2pt) node[below] {$\tau_1$};
			\begin{scope}
				\clip (100pt,0) circle(2pt) ;
				\fill[black] (100pt,-2pt) rectangle (98pt,2pt);
				\fill[white] (100pt,-2pt) rectangle (102pt,2pt);
			\end{scope}
			\draw (100pt,0) circle (2pt) node[below] {$\tau_2$};
			\draw[thick] (0,0) -- (100pt,0) node[pos=0.5, above] {$s$};
			\draw[dashed] (-30pt,50pt) to (0,0);
			\filldraw[color=black,fill=white,xshift=-2pt,yshift=-2pt] (-30pt,50pt) rectangle ++(4pt,4pt);
			\draw[dashed] (30pt,50pt) to (0,0);
			\filldraw[color=black,fill=white,xshift=-2pt,yshift=-2pt] (30pt,50pt) rectangle ++(4pt,4pt);
			\draw[dashed] (70pt,50pt) to (100pt,0);
			\filldraw[color=black,fill=white,xshift=-2pt,yshift=-2pt] (70pt,50pt) rectangle ++(4pt,4pt);
			\draw[dashed] (130pt,50pt) to (100pt,0);
			\filldraw[color=black,fill=white,xshift=-2pt,yshift=-2pt] (130pt,50pt) rectangle ++(4pt,4pt);
			\node[black] at (-25pt,25pt) {$\k_1$};
			\node[black] at (25pt,25pt) {$\k_2$};
			\node[black] at (75pt,25pt) {$\k_3$};
			\node[black] at (125pt,25pt) {$\k_4$};
		\end{tikzpicture}
		\label{single_exchange_diagram}
	\end{equation}
	where the exchanged field has mass parameter $\mu$ and each vertex carries two conformally coupled legs, so that $E_1=k_1+k_2$, $E_2=k_3+k_4$. Since $C_i=2$ and $M_i=1$, we
	have $N_1=N_2=0$: no differential operator is needed. Using \eqref{def_rescaled_diag} and \eqref{general_lambda_diagram}, the correlator \eqref{single_exchange_diagram} can be written as
	\begin{equation}
		\mathcal{F}=\frac{g^2\, H^4 \,\tau_0^4}{16 k_1 k_2 k_3 k_4} s  \times \Re \left(\tilde{\mathcal{F}}_{++}+\tilde{\mathcal{F}}_{+-} \right) 
		\label{correlator-single-exchange}
	\end{equation}
	where
	\begin{equation}
		\tilde{\mathcal{F}}_{+-}\left(E_1,E_2,s\right)
		=\int_1^\infty \d \lambda \int_1^\infty \d \lambdat \,P_{i\mu-1/2}(\lambda)\,P_{i\mu-1/2}(\lambdat)\,
		\frac{1}{E_1+s\lambda-i \epsilon}\frac{1}{E_2+s\lambdat+i \epsilon}    
	\end{equation}
	and
	\begin{equation}
		\begin{aligned}
			\tilde{\mathcal{F}}_{++}\left(E_1,E_2,s\right)
			=&-\int_1^\infty \d \lambda\int_1^\infty \d \lambdat \,P_{i\mu-1/2}(\lambda)\,P_{i\mu-1/2}(\lambdat) \label{tilde_F++_full} \\
			& \times\frac{1}{E_1+E_2+s (\lambda-\lambdat)-i \epsilon}
			\left(\frac{1}{E_1+s\lambda-i \epsilon} + \frac{1}{E_2+s\lambda-i \epsilon}\right)
			\,. 
		\end{aligned}
	\end{equation}
	In the second term of the sum in \eqref{tilde_F++_full}, we exchanged the integration variables $\lambda\leftrightarrow\lambdat$, leading to the above factorisation. Using the dispersive integral, for $z$ away from the cut of the Legendre function at $(-\infty,-1)$:
	\begin{equation}\label{eq: disperive integral lambda}
		\int_1^\infty \d\lambda\, \frac{P_{i\mu-1/2}(\lambda)}{\lambda+z}=  \frac{\pi}{\cosh(\pi \mu)} P_{i\mu-1/2}(z)\,,
	\end{equation}
	the factorised component $\tilde{\mathcal{F}}_{+-}$ reads, after taking the limit $\epsilon \to 0$:
	\begin{equation}
		\tilde{\mathcal{F}}_{+-}\left(E_1,E_2,s\right)
		=\frac{\pi^2}{s^2\cosh^2(\pi \mu)} \,
		P_{i\mu-1/2}(\lambda_u)\,P_{i\mu-1/2}(\lambda_v)\,,
		\label{tilde_F+-_2}
	\end{equation}
	where we defined $\lambda_u=E_1/s$ and $\lambda_v=E_2/s$, the inverse of the kinematic variables $u$ and $v$ often used in the cosmological bootstrap. Using the same dispersive integral, one can perform the first layer of integration in $\lambdat$ in \eqref{tilde_F++_full} to write
	\begin{equation}
		\begin{aligned}
			\tilde{\mathcal{F}}_{++}\left(E_1,E_2,s\right)
			&= 
			\frac{\pi}{s^2 \cosh(\pi \mu)}
			\int_1^\infty \d \lambda \,P_{i\mu-1/2}(\lambda) P_{i\mu-1/2}\left(-\lambda-(\lambda_u+\lambda_v)+i \epsilon\right)\\
			&\qquad \qquad \times\left(\frac{1}{\lambda+\lambda_u-i \epsilon} + \frac{1}{\lambda+\lambda_v-i \epsilon}\right)\,.
		\end{aligned}
		\label{F++_2}
	\end{equation}
	As anticipated, the Laplace-space avatar of the Schwinger-Keldysh contour prescriptions, i.e.~the complex deformations of the Laplace parameters (here $\lambda_{u,v} \to\lambda_{u,v}-i \epsilon$), removes any ambiguity: the Legendre function is never evaluated on its cut. The $i\epsilon$'s attached to the rational factors can then be set to zero, since the poles at $\lambda=-\lambda_{u,v}$ lie away from the integration range. We can recast the real part of the Laplace-form \eqref{F++_2}, which enters the correlator \eqref{correlator-single-exchange}, in a manifestly real form. For this, we use the connection formula, for $x>1$:
    \begin{equation}\label{eq: Legendre connection cut}
		P_{i\mu-1/2}(-x+i \epsilon)
		= \frac{2}{\pi}\cosh(\pi\mu)\,\Re\,Q_{i\mu-1/2}(x)
		\;-\; i\,\cosh(\pi\mu)\,P_{i\mu-1/2}(x)\,.
	\end{equation}
	Inserting \eqref{eq: Legendre connection cut} into \eqref{F++_2}, and
	taking into account that $P_{i\mu-1/2}$ is real on $(1,\infty)$, one obtains an expression with a manifestly real integrand, free from any $i\epsilon$,
	\begin{equation}\label{eq: F++ real}
		s^2\,\Re\tilde{\mathcal{F}}_{++}
		= 2\int_1^\infty \d\lambda\;P_{i\mu-1/2}(\lambda)\,
		\Re\, Q_{i\mu-1/2}(\lambda+\lambda_T)
		\left[\frac{1}{\lambda+\lambda_u}+\frac{1}{\lambda+\lambda_v}\right],
	\end{equation}
	where we introduced the total-energy variable $\lambda_T\equiv\lambda_u+\lambda_v=(E_1+E_2)/s$.
	
	We now show that the Laplace-space representations \eqref{tilde_F++_full} and \eqref{F++_2} (or equivalently \eqref{eq: F++ real}) give direct access both to the singularity structure of the correlator in the complexified energies, and to its complete analytic evaluation as a rapidly convergent series.

\paragraph{The total-energy singularity.}
A well known property of cosmological correlators is their non-analytic behaviour at vanishing total energy, here $E_1+E_2\to0$ at fixed $s$, i.e.~$\lambda_T\to0$. This lies outside the physical region $\lambda_{u,v}\geq1$ and is approached by analytic continuation, $\lambda_v\to-\lambda_u$, with $|\lambda_u|<1$ say. In the time-domain representation, the singularity arises from the early-time region where all vertices are simultaneously pushed to $\tau\to-\infty$, and its coefficient is the flat-space scattering amplitude \cite{Maldacena:2011nz,Raju:2012zr}.
The Laplace-space representation offers a precise dual picture of this mechanism, as we will show from \eqref{tilde_F++_full}. In this expression, the total energy enters through the single denominator $E_1+E_2+s(\lambda-\lambdat) -i\epsilon=s\left(\lambda_T+\lambda-\lambdat\right)-i\epsilon$, which is the only contribution to the integration layer over $\lambdat$ (up to the usual Legendre $P$ kernel). 
Then, to extract non-analyticities in $\tilde{\mathcal{F}}_{++}$ in the limit $\lambda_T\to 0$, we look for configurations where the integral over $\lambdat$ diverges, otherwise \eqref{tilde_F++_full} is manifestly analytic.
In that extent, we first note that the integrand in $\lambdat$ has a pole slightly shifted from the integration domain by the $i\epsilon$ factor.
Then, in the $\epsilon\to 0$ limit, the integral over $\lambdat$ can be rewritten as a sum of its Cauchy principal value and an imaginary part. Since we are interested in $\Re\,\tilde{\mathcal{F}}_{++}$, we stick to the principal value, and considering it as a function of $\lambda+\lambda_{T}$, the integral diverges for $\lambda+\lambda_{T}\to 1$ (because the pole in the integrand is located at the boundary of the integration domain, hence not regulated by the principal value). We obtain
\begin{equation}\label{eq_PV_log}
\mathrm{PV}\!\int_1^\infty \d\lambdat\,\frac{P_{i\mu-1/2}(\lambdat)}{\lambda+\lambda_T-\lambdat}
\underset{\lambda+\lambda_{T}\to 1}{=} \log\!\left(\lambda+\lambda_T-1\right)
+O(1)\,,
\end{equation}
where the finite terms are regular and analytic in $\lambda_T$. The location of this divergence at $\lambda+\lambda_T=1$ combined with the limit of interest $\lambda_{T}\to 0$ leads us to focus on the behaviour near $\lambda=1$ in the corresponding integral in \eqref{tilde_F++_full}. To do so, we split this integral into the two domains $\lambda\in $ $[1,1+\delta]$ and $[1+\delta,\infty]$, with $\delta>0$. We can focus on the first one, since the non-analyticities only come from the bound $\lambda=1$.
Using the decomposition \eqref{eq_PV_log} in terms of analytic and non-analytic pieces, and keeping the leading term in the Taylor expansion around $\lambda=1$ of the remaining $\lambda$ integrand\footnote{Considering the subleading terms in the Taylor expansion only give analytic contributions or subleading non-analyticities, i.e. $\propto \lambda_T^2 \log\!\left(\lambda_T\right)$}, this gives
\begin{equation}
    s^2\,\Re\tilde{\mathcal{F}}_{++}=\left(\frac{1}{1+\lambda_u}+\frac{1}{1+\lambda_v}\right)\int_1^{1+\delta}
    \d\lambda
    \log\!\left(\lambda+\lambda_T-1\right)
    +\mathrm{analytic}\,,
\end{equation}
which immediately leads, up to additional analytic contributions, to
\begin{equation}
\begin{aligned}
    s^2\,\Re\tilde{\mathcal{F}}_{++}
    &=\left[\frac{1}{1+\lambda_u}+\frac{1}{1+\lambda_v}\right]\lambda_T\log\lambda_T+\mathrm{analytic} \\
   & \;\xrightarrow[\lambda_v\to-\lambda_u]{}\;
    \frac{2}{1-\lambda_u^2}\,\lambda_T\log\lambda_T+\mathrm{analytic}\,.
    \end{aligned}
     \label{eq: total energy expansion}
\end{equation}
Here, two well-known features of the total-energy singularity are made manifest. First, the coefficient is independent of the mass of the exchanged field, as the Legendre weight only enters through $P_{i\mu-1/2}(1)=1$. Second, it is proportional to the flat-space amplitude for the exchange of a massless field, $\mathcal{A}_{\rm flat}=\left[E_1^2-s^2\right]^{-1}=s^{-2}(\lambda_u^2-1)^{-1}$, evaluated on the total-energy surface, recovering the general statement that masses become negligible in the early-time regime probed by this singularity. The mechanism transparently generalises to more complicated diagrams: the total-energy denominator vanishes at the Bunch-Davies corner of the dual integration region where all the $\lambda$'s equal unity, every Legendre weight trivialises there, and only the flat-space rational structure of the integrand survives. This is the Laplace-space avatar of all time integrals reaching the far past. Note finally that the coefficient in \eqref{eq: total energy expansion} blows up as $\lambda_u\to1$, i.e.~$\lambda_v\to-1$: this signals the collision of the total-energy singularity with a partial-energy singularity, to which we now turn.

	\paragraph{Partial-energy singularities.}
	Correlators are also singular when the energy flowing into a single vertex vanishes, here e.g.~$E_1+s\to0$, i.e.~$\lambda_u\to-1$, at fixed $\lambda_v\geq1$. As with the total-energy singularity, the Laplace-space way of seeing it still relies on the analysis of integral divergences, but with the roles of the two factors exchanged. In \eqref{tilde_F++_full}, it is now the rational factor $1/(\lambda_u+\lambda)$ whose pole at $\lambda=-\lambda_u$ reaches the integration domain boundary $\lambda=1$ as $\lambda_u\to-1$. Extracting the non-analytic pieces in this limit gives us the logarithm $-\log(1+\lambda_u)$. Performing the second integration layer with the remaining terms fixed at $\lambda_u=-1$, by using the dispersive integral over $\lambdat$, one finds:
    \begin{equation}
        \int_1^\infty\d\lambdat\,\frac{P_{i\mu-1/2}(\lambdat)}{1+\lambda_T-\lambdat-i\epsilon}=-\frac{\pi}{\cosh(\pi\mu)}P_{i\mu-1/2}\!\left(-(1+\lambda_T)+i\epsilon\right)\,,
    \end{equation}
    and then
    \begin{equation}\label{eq: partial energy expansion}
		s^2\,\tilde{\mathcal{F}}_{++}
		\;\xrightarrow[\;\lambda_u\to-1\;]{}\;
		-\frac{\pi}{\cosh(\pi\mu)}\,P_{i\mu-1/2}\!\left(e^{+i\pi}\lambda_v\right)\,\log\left(1+\lambda_u\right)+\mathrm{regular}\,.
	\end{equation}
Its real part, $-2\,\Re\, Q_{i\mu-1/2}(\lambda_v)\log(1+\lambda_u)$ according to Eq.~\eqref{eq: Legendre connection cut}, agrees with the form deduced directly from \eqref{eq: F++ real}, where the same mechanism is at play: the rational pole goes to the Bunch-Davies endpoint $\lambda=1$, and the residue is the value $\Re \,Q_{i\mu-1/2}(1+\lambda_T)\to\Re \,Q_{i\mu-1/2}(\lambda_v)$ of the other factor there. The origin of the singularity is clear: it arises when the pole of the partial energy denominator reaches the boundary of the integration domain, and the residue is the mass-dependent Legendre kernel itself---the Laplace-space expression of the three-point function of two conformally coupled and one massive field, with deformed arguments---in agreement with the factorisation of correlators into a product of a lower-point correlator and a lower-point amplitude near partial-energy singularities (see e.g.~\cite{    Baumann:2020dch,Goodhew:2020hob,Jazayeri:2021fvk}). For the factorised component \eqref{tilde_F+-_2}, the same behaviour is immediate: the logarithmic branch point of $P_{i\mu-1/2}(\lambda_u)$ at $\lambda_u=-1$ directly yields $s^2\tilde{\mathcal{F}}_{+-}\to-\frac{\pi}{\cosh(\pi\mu)}P_{i\mu-1/2}(\lambda_v)\log(1+\lambda_u)$, the residue being the three-point function at undeformed arguments.

\paragraph{The master series.}
We now turn to the complete evaluation of \eqref{eq: F++ real}. Two expansions reduce its integrand to a sum of elementary blocks. First, the large-argument hypergeometric representation of the Legendre $Q$ function (see \S14.3 in \cite{NIST:DLMF}),
\begin{equation}\label{eq: Q large argument}
    Q_{i\mu-1/2}(z)=\sqrt{\pi}\,\frac{\Gamma(\tfrac12+i\mu)}{\Gamma(1+i\mu)}\,(2z)^{-\frac12-i\mu}
    \sum_{n=0}^{\infty}a_n\,z^{-2n}\,,
    \qquad
    a_n=\frac{\left(\tfrac14+\tfrac{i\mu}{2}\right)_n\left(\tfrac34+\tfrac{i\mu}{2}\right)_n}{\left(1+i\mu\right)_n\,n!}\,,
\end{equation}
convergent for $|z|>1$ and hence on the whole range $\lambda+\lambda_T\geq 3$. Second, the rational factors expanded around $\lambda+\lambda_T$:
\begin{equation}\label{eq: rational expansion}
    \frac{1}{\lambda+\lambda_u}=\sum_{m=0}^{\infty}\frac{\lambda_v^{\,m}}{(\lambda+\lambda_T)^{m+1}}\,,
\end{equation}
with ratio $\lambda_v/(\lambda+\lambda_T)<1$ everywhere on the integration range, and similarly with $\lambda_u\leftrightarrow \lambda_v$. Each resulting term is then an integral of the Legendre function $P_{i\mu-1/2}$ against an inverse power $(\lambda+\lambda_T)^{-\rho}$, a one-parameter extension of the dispersive integral \eqref{eq: disperive integral lambda} that we now evaluate in closed form. Introducing a Schwinger parameter to write $(\lambda+\lambda_T)^{-\rho}=\Gamma(\rho)^{-1}\int_0^\infty\d t\;t^{\rho-1}e^{-t(\lambda+\lambda_T)}$, the $\lambda$ integral produces $\int_1^\infty\d\lambda\,e^{-t\lambda}P_{i\mu-1/2}(\lambda)=t^{-1}W_{0,i\mu}(2t)$, which is nothing else than the evaluation of the plane-wave representation \eqref{eq: massive mode plane wave} at $z=-it$, where the Hankel function turns into the Macdonald function, see Appendix~\ref{app: Whittaker func}. The remaining $t$ integral is then precisely the closed-form Laplace transform \eqref{eq: Whittaker Laplace closed} of the twisted massive mode function of Appendix~\ref{app:Whittaker}, at zero chemical potential and twist $\alpha=\rho-2$. We thus obtain \begin{equation}\label{eq: generalised dispersive integral}
T_\rho(\lambda_T)\equiv\int_1^\infty\d\lambda\;
\frac{P_{i\mu-1/2}(\lambda)}{(\lambda+\lambda_T)^{\rho}}
=\frac{2^{1-\rho}\,\Gamma\!\left(\rho-\tfrac12\pm i\mu\right)}{\Gamma(\rho)}\;
{}_2\tilde F_1\!\left(^{\rho-\frac12+i\mu,\ \rho-\frac12-i\mu}_{\qquad\quad \rho};\tfrac{1-\lambda_T}{2}\right),
\end{equation}
valid for $\Re\rho>\tfrac12$ and $\lambda_T$ away from the cut $(-\infty,-1]$. At $\rho=1$, using $\Gamma(\tfrac12+i\mu)\Gamma(\tfrac12-i\mu)=\pi/\cosh(\pi\mu)$ and \eqref{eq: Legendre def}, it reduces to \eqref{eq: disperive integral lambda}. Note that all parameters of the Legendre kernel \eqref{eq: Legendre def} are rigidly shifted by $\rho-1$: the deformation stays within the family of Laplace-space kernels of Appendix~\ref{app: 2F1}. Each term of the expansion of \eqref{eq: F++ real} is then a $T_\rho$ with $\rho=2n+m+\tfrac32+i\mu$, for which one of the two Gamma factors in \eqref{eq: generalised dispersive integral} reduces to a factorial, $\Gamma(\rho-\tfrac12-i\mu)=(2n+m)!$\,. We thus obtain the explicit result:
\begin{equation}\label{eq: F++ master series}
\boxed{\;
s^2\,\Re\tilde{\mathcal{F}}_{++}
= 2\sqrt{\pi}\;\Re\!\left[\frac{\Gamma(\tfrac12+i\mu)}{\Gamma(1+i\mu)}\,2^{-\frac12-i\mu}
\sum_{n,m=0}^{\infty}a_n\left(\lambda_u^{\,m}+\lambda_v^{\,m}\right)
T_{2n+m+\frac32+i\mu}(\lambda_T)\right].\;}
\end{equation}
As a consistency check, we show in Appendix~\ref{app: total energy series} how the total- and partial-energy behaviours \eqref{eq: total energy expansion} and \eqref{eq: partial energy expansion}, manifest at the level of the integral representations, are also encoded in the series \eqref{eq: F++ master series}, as collective effects of its towers.

	\paragraph{Convergence and structure.}
	The series converges exponentially. From the bound $|T_\rho(\lambda_T)|\leq \max_{[1,\infty)}|P_{i\mu-1/2}|\,(1+\lambda_T)^{1-\Re\rho}/(\Re\rho-1)$, the $m$ sum is dominated by a geometric series of ratio $\max(\lambda_u,\lambda_v)/(1+\lambda_T)<1$, and the $n$ sum by a geometric series of ratio $(1+\lambda_T)^{-2}\leq 1/9$. Since $\rho-\tfrac32-i\mu=2n+m$, the natural truncation parameter is the total order $N$, keeping all terms with $2n+m\leq N$, each unit of which costs one inverse power of $1+\lambda_T$. The resulting error decays geometrically, as $\left[\max(\lambda_u,\lambda_v)/(1+\lambda_T)\right]^{N}$, in quantitative agreement with the above bounds.

\begin{figure}[t]
		\centering
		\includegraphics[width=\textwidth]{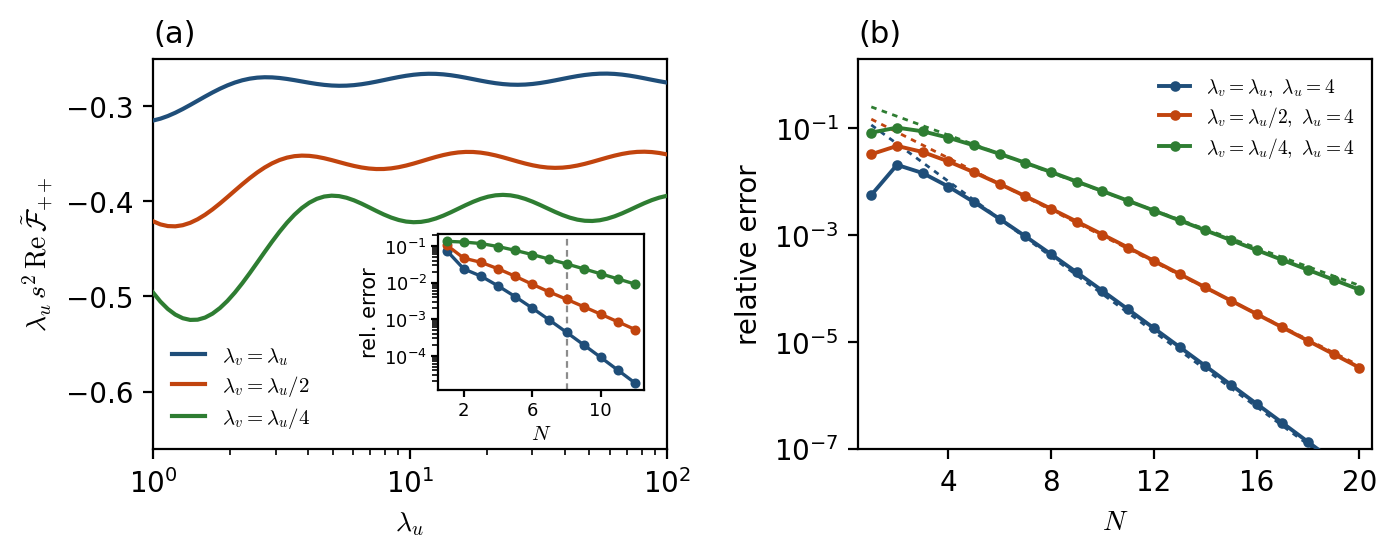}
		\caption{Numerical performance of the master series \eqref{eq: F++ master series}, truncated at total order $2n+m\leq N$, for $\mu=2$.
			\textbf{(a)} The single-exchange correlator $s^2\,\Re\tilde{\mathcal{F}}_{++}$ rescaled by $\lambda_u$, exposing the oscillatory cosmological collider signal on top of the smooth effective-field-theory background, along three slices of fixed ratio $\lambda_v/\lambda_u\in\{1,\tfrac12,\tfrac14\}$; the curves are the series truncated at $N=8$ ($25$ terms). Manifestly symmetric under $\lambda_u\leftrightarrow\lambda_v$, this single expansion covers the whole domain at once, including the equal-energy configuration $\lambda_v=\lambda_u$ that sits at the boundary between the orderings handled by separate expansions in other approaches. \emph{Inset}: the relative error of the truncation, maximised over each slice, falls geometrically with $N$ (dashed line: the order $N=8$ of the main panel).
			\textbf{(b)} The same relative error at fixed kinematics $\lambda_u=4$, for the three ratios of panel (a), pushed to $N=20$. Dashed lines show the geometric estimate $\propto\left[\max(\lambda_u,\lambda_v)/(1+\lambda_T)\right]^{N}$, anchored in the asymptotic regime to expose its slope: the decay is geometric throughout, fastest at equal energy and slowing towards the hierarchical configuration, in quantitative agreement with the predicted rate.}
		\label{fig: single exchange convergence}
	\end{figure}

	Several structural features deserve emphasis. 
    First, our simple representation readily gives the full correlator, treating on the same footing both the effective field theory background and the cosmological collider oscillatory signal \cite{Chen:2009zp,Chen:2009we,Noumi:2012vr,Arkani-Hamed:2015bza}, well visible in Fig.~\ref{fig: single exchange convergence}.
    Second, the series \eqref{eq: F++ master series} is resummed entirely in the ratios $\lambda_{u,v}/(1+\lambda_T)$ and is manifestly symmetric under $\lambda_u\leftrightarrow\lambda_v$: a single expansion converges geometrically across the entire kinematic domain, with no patching of the two orderings $\lambda_u<\lambda_v$ and $\lambda_u>\lambda_v$ that earlier approaches handle through separate, asymmetric expansions in $\lambda_u$ or $\lambda_v$~\cite{Arkani-Hamed:2018kmz,Qin:2022fbv,Werth:2024mjg}. Third, the equal-energy configuration $\lambda_u=\lambda_v$, which lies precisely at the boundary between these two regions, is here treated like any other and is in fact where convergence is fastest, the rate decreasing towards the hierarchical configurations $\lambda_u\gg\lambda_v$. This behaviour is illustrated in Fig.~\ref{fig: single exchange convergence}: already at $N=8$, i.e.~with $25$ terms, the truncated series is indistinguishable from a direct numerical integration, and a few more orders reach machine precision. Note that the fastest-converging regular configurations fill the bulk of the observationally relevant kinematic space, while the less rapidly-converging hierarchical corner is precisely where the analytically known cosmological collider signal dominates.

	\subsection{Generalisations} 
	\label{sec:generalisations}
	
	The construction above was carried out in detail for a definite setup---cosmological correlators and polynomial interactions of a
	conformally coupled field $\varphi$ with massive scalars $\sigma^A$ in de Sitter---but its
	guiding idea, trading every non-trivial time dependence for plane waves through a Laplace
	representation, reaches well beyond that case. We list here, in increasing order of departure
	from the explicit framework, the situations to which the method applies essentially unchanged.

	\paragraph{Wavefunction.} We focused on cosmological correlators but the same construction goes through for wavefunction coefficients. The only differences are that there is no Schwinger-Keldysh index in that context, and that the bulk-to-bulk propagator has an additional component, which can however be treated analogously. We describe these changes and summarise the corresponding rules in Appendix~\ref{app:WFU}.

	\paragraph{Derivative interactions.} Adding spatial or temporal derivatives to the vertices
	changes nothing essential. A spatial gradient $a^{-1}\partial_{x^i}$ only produces kinematic factors and adds one to the conformal time power, so one remains within the
	``plane-waves only'' analysis of Section~\ref{diag_rules_correlators}. For temporal derivatives, the only change compared with polynomial interactions is that a field $\sigma$ is replaced by $a^{-1}\partial_\eta\sigma$. Hence, differentiating the plane-wave
	representation of the mode function~\eqref{Borel_rep_massive} preserves the Laplace structure and keeps the conformal time power positive, simply bringing down a factor of the dual variable.\footnote{This is equivalent to expressing the time derivative of the mode function as a combination of Hankel functions and
		apply~\eqref{Borel_rep_massive} to each.}
	
	\paragraph{Reduced sound speeds.} Suppose each field propagates with its own reduced sound speed
	$c_s$. Up to overall multiplicative factors built from these sound speeds---the analogue of the
	prefactor relating $\mathcal{F}$ to $\tilde{\mathcal{F}}_{\{a_i\}}$ in~\eqref{def_rescaled_diag}---the
	formalism applies unchanged once all the external energies $E_i$ and internal momenta $s_{ij}$ are
	rescaled by the appropriate sound speeds. The reason is that the Laplace representation of each
	mode function, in particular~\eqref{eq: mode function dispersive 1}, goes through identically with
	$z=c_s k\tau$; inserting it into the time integrals, the only net effect is precisely this
	rescaling of $E_i$ and $s_{ij}$.
	
	\paragraph{Spin-1 field, chemical potential and time-dependent couplings.} Still considering de Sitter space, the transverse helicity states of a spin-1 field with a chemical potential have a mode function proportional to a Whittaker function
	(Appendix~\ref{app:Whittaker}). The building block entering the relevant seed diagrams is then
	$W_{i\tk,i\mu}(2iz)/z$, with no other time dependence, see e.g.~\cite{Qin:2022fbv,Qin:2023ejc,Qin:2025xct}, and its plane-wave representation is given
	in~\eqref{eq: helical Legendre}, so the method applies directly. Time-dependent coupling constants
	can likewise be accommodated: by Fourier transform, they reduce to oscillatory factors
	$\cos(\omega t)$ in cosmic time, i.e.~to powers $\tau^{\pm i \omega/H}$ in
	conformal time. The general object to consider is therefore $z^\alpha W_{i\tk,i\mu}(2iz)$ with arbitrary
	complex $\alpha$. We use the Laplace method for this twisted Whittaker mode function in Appendix~\ref{app:Whittaker}, leading to the
	plane-wave representation~\eqref{eq: twisted Whittaker master}. The latter holds only for
	$\Re(\alpha)<0$, but this is no obstruction, exactly as for the $N_i>0$ case of the
	``plane-waves only'' discussion in Section~\ref{diag_rules_correlators}: when $\Re(\alpha)\geq0$ one chooses an integer $n$ such that
	$\alpha'=\alpha-n$ satisfies $\Re(\alpha')<0$, applies the representation to $\alpha'$, and
	recovers the desired power $z^\alpha=z^{\alpha'+n}$ by differentiating $n$ times with respect to
	the external energy $E_i$.
	
	\paragraph{Beyond de Sitter.} Finally, in a generic situation, in particular beyond de
	Sitter space, the same logic still applies, provided one can compute the Laplace transform of the corresponding bulk function, which can always be done numerically. The flat space structure, i.e.~the $\mathcal{V}_i$ in \eqref{general_lambda_diagram}, remains universal, and the specificity of the model, as above, is in the Laplace-space kernel.

	\section{Conclusions and Outlook}
	
	In this work we have introduced a Laplace-space approach to cosmological correlators, built on the
	simple fact that, deep inside the horizon, every mode oscillates as a flat-space plane wave. Laplace-transforming the bulk time dependence in the conformal variable exposes the analytic structure imprinted by the early-time, Bunch-Davies behaviour of the mode: a branch cut whose discontinuity, the kernel that dresses the flat-space content back into the curved-space one, resolves each curved-space mode function into a continuous superposition of plane waves labelled by a dual variable $\lambda$. We have shown that the same representation can be reached from a complementary direction for massive scalars, by Borel resumming the early-time asymptotic series of the mode function, although the two methods give distinct integral representations in more general situations.
	
	With the Laplace representation at hand, the nested time integrals of the in-in formalism collapse: each
	massive internal line becomes a plane wave, every bulk time integral reduces to an elementary
	flat-space one, and all the information about the spacetime geometry, field content and dynamics is
	carried by the known $\lambda$-space kernel. We have summarised this into a set of diagrammatic rules
	that produce, directly from a diagram, the Laplace-space integrand for de Sitter correlators with
	conformally coupled external legs and massive internal exchanges. The difficulty of the curved-space problem does not disappear but is relocated: the entire departure from flat space is repackaged into a single rigid structure, the known Laplace-space kernels, each fixed by a dual equation of motion and integrated over the fixed domain $\lambda\in[1,\infty)$ against flat-space rational denominators. Applied to the massive single-exchange correlator, this representation makes the total- and partial-energy singularities transparent ``from flat space'' and yields a single closed-form, very rapidly convergent series valid throughout the entire kinematic domain, treating on equal footing the effective field theory background and the cosmological collider signal. This demonstrates that the Laplace-space approach is not merely elegant but also powerful.

	Several directions open up naturally. On the computational side, the diagrammatic rules apply to
	more intricate diagrams, at both tree and loop level, radiative corrections reducing to loop-momentum integrals over flat-space quantities dressed by the Laplace-space kernels, as well as to the various extensions laid out in
	Section~\ref{sec:generalisations}: wavefunction coefficients, de Sitter-breaking setups with derivative interactions, non-trivial
	sound speeds and time-dependent couplings. We have already shown how our Laplace representation applies to transverse modes of a spin-1 field endowed with a chemical potential, and it would be interesting to develop a systematic treatment of
	spinning fields. Should the property of a single representation valid in all kinematics hold for more general correlators, this would also be very valuable in order to confront theory with data.

	Beyond these calculational gains, where the strength of the method is already manifest, the
	genuinely ``from flat space'' character of the construction points to a deeper payoff. It is most
	visible in the way the total- and partial-energy singularities of the correlator are inherited from
	flat space, but its scope is broader. The emergence of cosmological correlators from flat-space data has so far been explored mostly for massless or conformally coupled fields, notably by dressing flat-space scattering amplitudes~\cite{Chowdhury:2023arc,Chowdhury:2025ohm,Chowdhury:2026upp,Das:2025qsh,Das:2026vfv,Ansari:2026xkm,Ansari:2026sjf}. Our construction is from flat space but in a different sense: its flat-space building blocks are massless correlators rather than amplitudes, and it covers the massive exchanges, and essentially any theory of interest in primordial cosmology. It thus offers a new perspective on conceptual questions such as the analytic structure of cosmological correlators and the imprint of unitarity carried by their discontinuities, suggesting a way to elucidate the former by uplifting that of their comparatively simpler flat-space counterparts. The reach may run further still, beyond cosmological correlators altogether: wherever a curved background becomes asymptotically flat, the modes revert to ordinary Minkowski plane waves, and a Laplace construction of the same kind should apply, around black holes for instance, with the radial coordinate playing the role of the conformal time. Resting on so elementary a fact and yet so general, the Laplace approach is at once a calculational engine and a new route to the structure of curved-space observables.

	\paragraph{Acknowledgements.} We thank Denis Werth for initial collaboration and Guillaume Faye, Sebastian Garcia-Saenz, Austin Joyce and Zhong-Zhi Xianyu for useful discussions. We are also grateful for the feedback of the many participants in several scientific events where Nathan Belrhali presented this work while it was in preparation: the program 
	\href{https://indico.ijclab.in2p3.fr/event/11373/}{CoBALt} held at the Institut Pascal at Universit\'e Paris-Saclay with the support of the program ``Investissements
	d'avenir'' ANR-11-IDEX-0003-01, the \href{https://indico.in2p3.fr/event/35602/}{TUG workshop} at IPhT Saclay, the 41st annual IAP symposium \href{https://indico.iap.fr/event/36/}{Inflation 2025}, the \href{https://earlyuniverse.discussingresearch.com/}{Early Universe from Home} 2026 online conference, and the \href{https://indico.cern.ch/event/1556583/overview}{PONT 2026} conference in Avignon.

	\appendix
	\section{Special functions}\label{app: special functions}
	This appendix collects the special functions used in the main text and in
	Appendix~\ref{app:Whittaker}, together with the properties we use. The mode functions in conformal time are
	\emph{confluent} hypergeometric functions---Whittaker functions $W_{\kappa,\nu}$, of which the
	de Sitter Hankel/modified-Bessel case is a reduction. The dual-space kernels produced by
	the Laplace transform are \emph{Gauss} hypergeometric functions ${}_2F_1$, of which Legendre,
	associated Legendre and Gegenbauer functions are named reductions. The Laplace transform thus
	trades a confluent parent in time for a Gauss parent in $\lambda$. We treat the two in turn.
	
	\subsection{Hankel and Whittaker functions}\label{app: Whittaker func}

	The Whittaker function $W_{\kappa,\nu}$ is the solution of the confluent hypergeometric equation
	\begin{equation}\label{eq: Whittaker equation app}
		W_{\kappa,\nu}''(x)+\left(-\frac14+\frac{\kappa}{x}+\frac{\tfrac14-\nu^2}{x^2}\right)W_{\kappa,\nu}(x)=0\,,
	\end{equation}
	that decays at large
	argument, with leading behaviour
	\begin{equation}\label{eq: Whittaker asymptotics app}
		W_{\kappa,\nu}(x)\;\underset{x\to+\infty}{\sim}\;x^{\kappa}\,e^{-x/2}\,.
	\end{equation}
	This asymptotic behaviour is the one relevant to implement the Bunch-Davies condition in the context of the helical mode
	function~\eqref{eq:mode-function-W}, built directly on $W_{i\tk,i\mu}$. For $\kappa =0$, the Whittaker function reduces to a modified
	Bessel (Macdonald) function, $W_{0,\nu}(z)=\sqrt{z/\pi}\,K_\nu(z/2)$, and further note that for $-3 \pi/2 < \arg(z)\leq 0$,
	one has $K_{i\mu}(iz)=\frac{i \pi}{2} e^{-\pi \mu/2} H^{(1)}_{i \mu}(-z)$, which ties it to the massive mode function~\eqref{eq: massive mode plane wave}.

	\subsection{The Gauss hypergeometric function and the Laplace-space kernels}\label{app: 2F1}
	The Gauss hypergeometric function is defined by the series
	\begin{equation}\label{eq: 2F1 def}
		{}_2F_1\left(a,b;c;z\right) = \sum_{n=0}^\infty\frac{(a)_n(b)_n}{(c)_n}\frac{z^n}{n!}\;,
		\qquad (a)_n = \frac{\Gamma(a+n)}{\Gamma(a)}\;,
	\end{equation}
	convergent for $|z|<1$ and continued to $z\in\mathbb{C}\setminus[1,\infty)$; it carries a
	power-law branch cut on $[1,\infty)$. We also use the regularised function
	${}_2\tilde{F}_1(a,b;c;z)={}_2F_1(a,b;c;z)/\Gamma(c)$, which is entire in the parameter $c$. The
	single property we need is its discontinuity across its cut, i.e.~for $z>1$ (\S15.8 in \cite{NIST:DLMF}):
	\begin{equation}\label{eq: disc 2F1 general}
		\Disc_z\left[{}_2\tilde{F}_1\left(a,b;c;z\right)\right]
		= 2\pi i\,\frac{z^{1-c}(z-1)^{c-a-b}}{\Gamma(a)\Gamma(b)}\;
		{}_2\tilde{F}_1\left(1-b,1-a;c-a-b+1;1-z\right)\,.
	\end{equation}
	
	Three named reductions of ${}_2F_1$ appear as Laplace-space kernels.
	\begin{description}[leftmargin=1.4em,itemsep=2pt,topsep=3pt]
		\item[Legendre.] The kernel of the de Sitter massive mode function is the Legendre function of
		degree $\nu=i\mu-\tfrac12$, with $\nu(\nu+1)=-(\mu^2+\tfrac14)$,
		\begin{equation}\label{eq: Legendre def}
			P_{i\mu-1/2}(\lambda)\equiv\,{}_2F_1\!\left(\tfrac12+i\mu,\,\tfrac12-i\mu;\,1;\,\tfrac{1-\lambda}{2}\right).
		\end{equation}
		It solves the Legendre equation~\eqref{eq: Legendre dual equation} and is the weight in the
		plane-wave representation~\eqref{eq: massive mode plane wave} and the diagrammatic rules of
		Section~\ref{diag_rules_correlators}. Its companion solution $Q_{i\mu-1/2}$ and the behaviour of
		$P_{i\mu-1/2}$ near $\lambda=-1$, used to select the physical solution, are discussed in
		Section~\ref{subsec: dual equation}.
		\item[Associated Legendre.] Turning on a chemical potential promotes the kernel to the
		associated Legendre function of order $m$ (here $m=i\tk$), which is the ${}_2F_1$
		of~\eqref{eq: Legendre def} dressed by a non-integer power,
		\begin{equation}\label{eq: assoc Legendre def}
			P^{m}_{i\mu-1/2}(\lambda)
			= \left(\frac{\lambda+1}{\lambda-1}\right)^{\!m/2}
			{}_2\tilde{F}_1\!\left(\tfrac12+i\mu,\,\tfrac12-i\mu;\,1-m;\,\tfrac{1-\lambda}{2}\right),
			\qquad \lambda>1\,.
		\end{equation}
		It is the kernel of the untwisted helical case, Eq.~\eqref{eq: helical Legendre}.
		\item[Gegenbauer.] A twist without chemical potential produces instead a Gegenbauer function,
		\begin{equation}\label{eq: Gegenbauer def}
			C^{(\gamma)}_{a}(\lambda)
			= \frac{\Gamma(a+2\gamma)}{\Gamma(2\gamma)\,\Gamma(a+1)}\,
			{}_2F_1\!\left(-a,\,a+2\gamma;\,\gamma+\tfrac12;\,\tfrac{1-\lambda}{2}\right),
		\end{equation}
		which appears with $\gamma=-\tfrac12-\alpha$ and $a=\tfrac12+\alpha-i\mu$ in the twisted-massive
		representation~\eqref{eq: twisted massive Gegenbauer}.
	\end{description}

	\section{Plane-wave decomposition for de Sitter-breaking mode functions}
	\label{app:Whittaker}
	The method of Section~\ref{sec: Laplace space} is not specific to the de Sitter massive mode
	function. It applies to any mode function, and  
when its equation of motion has polynomial coefficients, it is immediate to obtain the dual differential equation verified by its Laplace transform. We carry it out here for the most general case relevant to this work---a massive spin-1 field with a
	helical chemical potential, dressed by an arbitrary twist---and recover the chemical-potential
	and twisted-massive cases as limits. Throughout we use the notation of
	Section~\ref{sec: Laplace space}.
	
	\subsection{The twisted Whittaker mode function}
	The equation of motion for the mode function of transverse helicity states of a massive spin-1 field reads (see, e.g.~\cite{Jazayeri:2023kji})
	\begin{equation}
		\frac{d^2}{d \tau^2}\sigma^{\pm}_k+\big[k^2\pm 2 a k\kappa+a^2m^2\big]\sigma_k^{\pm}=0\,,
	\end{equation}
	whose Bunch-Davies solution in de Sitter space is a Whittaker function,
	\begin{equation}
		\sigma_k^\pm(\tau) = \frac{e^{-\tk\pi/2}}{\sqrt{2k}}\,W_{i\tk,i\mu}(2ik\tau)\,,
		\label{eq:mode-function-W}
	\end{equation}
	with $\mu^2\equiv m^2/H^2-\tfrac14$ and $\tk\equiv\pm\kappa/H$. As $\tk\to0$
	it reduces to the standard massive mode function,
	\begin{equation}
		\lim_{\tk\to0}\sigma^{\pm}_k(\tau)=\tfrac{\sqrt{\pi}}{2}\,e^{-\frac{\pi\mu}{2}+\frac{i\pi}{4}}(-\tau)^{\frac12}H_{i\mu}(-k\tau)\,.
	\end{equation}
	As in Section~\ref{sec: Laplace space}, we decompose not the bare mode function but the twisted
	object entering a bulk time integral: the mode function dressed by a power $z^\alpha$ whose
	exponent $\alpha$ encodes the conformal-time weight carried by a vertex. With $z=k\tau$ and up to normalisation, this is
	\begin{equation}\label{eq: twisted Whittaker bulk object}
		\F(z)=(iz)^\alpha\,W_{i\tk,i\mu}(2iz)\,.
	\end{equation}

	\subsection{Laplace transform and plane-wave representation}
	\paragraph{Dual equation.}
	Laplace-transforming the Wick-rotated Whittaker equation as in
	Section~\ref{subsec: dual equation}---after multiplying by $z^2$ to render its coefficients
	polynomial---trades the time evolution for a second-order equation in $\lambda$,
	\begin{equation}\label{eq: Whittaker dual equation}
		(\lambda^2-1)\,\hat{\F}'' + \big(2(\alpha+2)\lambda - 2i\tk\big)\hat{\F}'
		+ \big((\alpha+\tfrac32)^2+\mu^2\big)\hat{\F} = 0\,,
	\end{equation}
	which reduces to the Legendre equation~\eqref{eq: Legendre dual equation} of the main text at
	$\alpha=-1$, $\tk=0$. Its regular singular points sit at $\lambda=\pm1$; at the early-time end
	$\lambda=-1$ the indicial exponents are $\{0,\,-(\alpha+1)-i\tk\}$. As in
	Section~\ref{subsec: dual equation}, the Bunch-Davies behaviour selects and normalises the
	physical solution,
	\begin{equation}\label{eq: Whittaker Laplace closed}
		\mathcal{L}\!\left[z^\alpha W_{i\tk,i\mu}(2z)\right]\!(\lambda)
		= 2^{-\alpha-1}\,\Gamma\!\left(\tfrac32+\alpha\pm i\mu\right)\;
		{} _2\tilde{F}_1\!\left(^{\frac32+\alpha+i\mu,\ \frac32+\alpha-i\mu}_{\qquad 2+\alpha-i\tk};\tfrac{1-\lambda}{2}\right).
	\end{equation}
	Near $z=0$, the function $z^\alpha W_{i\tk,i\mu}(2z)$
	behaves as $z^{\alpha+1/2\pm i\mu}$, and hence the Laplace integral converges only for $\Re(\alpha)>-\tfrac32+|\Im(\mu)|$, but the right-hand side can be analytically continued beyond that strip, only having two series of poles at
	$\alpha^\pm_n=-\left(n+\tfrac32\pm i\mu\right)$, coming from the $\Gamma\!\left(\tfrac32+\alpha\pm i\mu\right)$
	factors. At $\alpha=-1$, $\tk=0$ the two indicial exponents coincide and the selection proceeds through the
	marginal logarithm of Section~\ref{subsec: dual equation}; a non-zero chemical potential lifts
	the degeneracy to $\{0,-i\tk\}$, deforming that logarithm into a non-integer power, as
	anticipated there. The same closed form follows equally by direct integration, using 7.621.3 in \cite{bibcite_894} and the Pfaff transformation 15.8.1 in \cite{NIST:DLMF}.
	
	\paragraph{Plane-wave representation.}
	The transform~\eqref{eq: Whittaker Laplace closed} is analytic for $\Re(\lambda)>-1$ and carries a branch cut on
	$\lambda\in(-\infty,-1)$, exactly as in Section~\ref{sec: Laplace space}. We read its
	discontinuity off the connection formula~\eqref{eq: disc 2F1 general} and invert through the
	dispersion relation~\eqref{eq: mode function dispersive 1}, closing the contour around the cut.
	As in the main text, the inversion drops the small arc $C_\epsilon$ encircling the branch point
	$\lambda=-1$; with the early-time exponent of~\eqref{eq: twisted Whittaker bulk object}, this arc
	scales as $\epsilon^{-\Re(\alpha)}$ and hence vanishes for $\Re(\alpha)<0$. Undoing the Wick
	rotation as in Section~\ref{sec: Laplace space}---rotating $z\to iz$, so that the real
	exponential becomes the plane wave $e^{-i\lambda z}$---then yields the plane-wave representation
	of the twisted Whittaker mode function,
	\begin{equation}\label{eq: twisted Whittaker master}
		\boxed{(iz)^\alpha W_{i\tk,i\mu}(2iz)=\int\displaylimits_1^\infty\!\d\lambda\;e^{-i\lambda z}
			\left(\frac{2}{\lambda^2-1}\right)^{\!\alpha+1}\!\left(\frac{\lambda+1}{\lambda-1}\right)^{\!i\tk}
			{}_2\tilde F_1\!\left(^{-\frac12-\alpha+i\mu,\ -\frac12-\alpha-i\mu}_{\qquad\ -\alpha-i\tk};\tfrac{1-\lambda}{2}\right)\!.}
	\end{equation}
	This is the most general representation used in this work. Although the underlying Laplace
	transform converges only in the strip $\Re(\alpha)>-\tfrac32+|\Im(\mu)|$, its analytic continuation has poles at $\alpha_n^{\pm}$, and its inversion requires
	$\Re(\alpha)<0$, the integral on the right-hand side converges for \emph{all} $\Re(\alpha)<0$, with
	no restriction on $\mu$: near $\lambda=1$ its kernel behaves as $(\lambda-1)^{-\alpha-1}$, with a
	$\mu$-independent exponent, integrable there whenever $\Re(\alpha)<0$. By analytic continuation in
	$\alpha$ and in $\mu$, the representation~\eqref{eq: twisted Whittaker master} therefore holds
	throughout $\Re(\alpha)<0$ for every physical mass.

	\paragraph{Special cases.} Two limits collapse the regularised hypergeometric kernel of~\eqref{eq: twisted Whittaker master}
	onto a named function. For the untwisted helical case $\alpha=-1$, corresponding to the seed interactions studied, e.g.\ in \cite{Qin:2023ejc,Qin:2025xct}, it becomes the associated
	Legendre function, giving the chemical-potential representation
	\begin{equation}\label{eq: helical Legendre}
		\frac{1}{iz}\,W_{i\tk,i\mu}(2iz)=\int\displaylimits_1^\infty\!\d\lambda\;e^{-i\lambda z}
		\left(\frac{\lambda+1}{\lambda-1}\right)^{\!i\tk/2}P^{i\tk}_{i\mu-\frac12}(\lambda)\,.
	\end{equation}
	For a vanishing chemical potential $\tk=0$ it reduces to a Gegenbauer function, giving the
	twisted-massive representation
	\begin{equation}\label{eq: twisted massive Gegenbauer}
		(iz)^\alpha W_{0,i\mu}(2iz)=\frac{\Gamma(-1-2\alpha)\,\Gamma\!\left(\tfrac32+\alpha-i\mu\right)}{\Gamma\!\left(-\tfrac12-\alpha-i\mu\right)\Gamma(-\alpha)}
		\int\displaylimits_1^\infty\!\d\lambda\;e^{-i\lambda z}\left(\frac{2}{\lambda^2-1}\right)^{\!\alpha+1}
		C^{\left(-\frac12-\alpha\right)}_{\frac12+\alpha-i\mu}(\lambda)\,,
	\end{equation}
	where $W_{0,i\mu}(2iz)=\sqrt{2iz/\pi}\,K_{i\mu}(iz)$.

	\subsection{Borel resummation and an alternative representation}
	\label{sec:Borel-Whittaker}
	
	The Laplace route of the previous subsection is not the only way to reach an integral
	representation of the twisted Whittaker mode function: as in Section~\ref{subsec_Borel}, the same
	object can be reconstructed by Borel-resumming its early-time expansion. Away from the de
	Sitter-invariant case, however, the two constructions no longer coincide. As anticipated at the
	end of Section~\ref{subsec_Borel}, the twist leaves an overall power of $z$ outside the dispersive
	integral, so the Borel sum returns a representation that is genuinely \emph{different} from---and
	less economical than---the master formula~\eqref{eq: twisted Whittaker master}. We make this
	explicit here.
	
	As in the main text we decompose the twisted bulk object~\eqref{eq: twisted Whittaker bulk object},
	$\F(z)=(iz)^\alpha W_{i\tk,i\mu}(2iz)$ with $z=s\tau$, whose early-time behaviour is fixed by the
	Whittaker asymptotics~\eqref{eq: Whittaker asymptotics app}, $\F(z)\sim(iz)^\alpha(2iz)^{i\tk}e^{-iz}$.
	To capture the corrections we insert the perturbative ansatz
	\begin{equation}
		W_{\kappa,\nu}(x) \sim x^\kappa e^{-\frac{x}{2}} \sum_{n=0}^{\infty} \frac{\beta_n(\kappa,\nu)}{x^n}\,,
		\qquad \beta_0=1\,,
		\label{asymptotic_series_W}
	\end{equation}
	into the Whittaker equation~\eqref{eq: Whittaker equation app}, which fixes the recursion
	\begin{equation}
		\ba
		\beta_n&=-\frac{(n-\kappa)(n-1-\kappa)+\frac14-\nu^2}{n}\,\beta_{n-1}\\
		&=-\frac{\left(\tfrac12+\nu+n-1-\kappa\right)\left(\tfrac12-\nu+n-1-\kappa\right)}{n}\,\beta_{n-1}\,,
		\ea
	\end{equation}
	with closed-form solution
	\begin{equation}
		\beta_n=(-1)^n\frac{\left(\tfrac12+\nu-\kappa\right)_n \left(\tfrac12-\nu-\kappa\right)_n}{n!}\,.
	\end{equation}
	Since $(a)_n\sim n!$ the coefficients grow factorially and the series is asymptotic, exactly as in
	Section~\ref{subsec_Borel}. Borel-resumming it with~\eqref{Borel_sum_generic} and specialising to
	$\kappa=i\tk$, $\nu=i\mu$, $x=2is\tau$, the twisted object becomes
	\begin{equation}
		\F(\tau,s)=(is\tau)^\alpha (2is\tau)^{i\tk}\,e^{-is\tau}\,(-s\tau)
		\int_0^\infty \d\lambda\,e^{\lambda s\tau}\,
		{}_2F_1\!\left(^{\frac12+i\mu-i\tk,\ \frac12-i\mu-i\tk}_{\qquad\qquad 1};-\tfrac{i\lambda}{2}\right)\,.
	\end{equation}
	For $\tau$ in the
	lower-half plane, $\Im(s\tau)<0$, deforming the contour onto the negative imaginary axis and shifting
	the integration variable as in Section~\ref{subsec_Borel}, we obtain
	\begin{equation}\label{eq: twisted Whittaker Borel}
		\boxed{(is\tau)^\alpha W_{i\tk,i\mu}(2is\tau)=2^{i\tk}(is\tau)^{1+\alpha+i\tk}
			\int\displaylimits_1^\infty\!\d\lambda\;e^{-is\lambda\tau}\,
			{}_2F_1\!\left(^{\frac12+i\mu-i\tk,\ \frac12-i\mu-i\tk}_{\qquad\qquad 1};\tfrac{1-\lambda}{2}\right).}
	\end{equation}
	
	Equation~\eqref{eq: twisted Whittaker Borel} is the alternative representation announced above. It
	reconstructs the same twisted mode function as the master formula~\eqref{eq: twisted Whittaker master},
	but organises it differently. The Borel route leaves the early-time related power
	$(is\tau)^{1+\alpha+i\tk}$ \emph{outside} the dispersive integral and weights the plane waves by the
	bare Gauss kernel ${}_2F_1\!\left(\cdots;1;\tfrac{1-\lambda}{2}\right)$, whose first two parameters
	are shifted by the chemical potential through $i\tk$. The Laplace construction, by contrast, absorbs
	the entire twist into the weight: nothing is left outside the integral, and the kernel is
	the regularised ${}_2\tilde F_1$ of~\eqref{eq: twisted Whittaker master}, carrying the twist in its
	third parameter and the chemical potential in the accompanying powers. For the dual-space
	manipulations of Section~\ref{sec: correlators}, where these integrals are performed against one
	another, the Laplace form is the more convenient of the two, which is why we adopt it as the master
	representation.
	
	The two representations meet in the de Sitter-invariant limit. At $\tk=0$ and $\alpha=-1$ the
	prefactor $2^{i\tk}(is\tau)^{1+\alpha+i\tk}$ reduces to unity and the kernel collapses onto the
	Legendre function~\eqref{eq: Legendre def}, so that~\eqref{eq: twisted Whittaker Borel} becomes
	$(is\tau)^{-1}W_{0,i\mu}(2is\tau)=\int_1^\infty\d\lambda\,e^{-is\lambda\tau}\,P_{i\mu-1/2}(\lambda)$,
	which is exactly the plane-wave representation~\eqref{eq: massive mode plane wave} of the massive
	mode function.

	\section{Correlator singularities from the master series}
	\label{app: total energy series}
	
	In the main text, the total- and partial-energy behaviours \eqref{eq: total energy expansion} and \eqref{eq: partial energy expansion} were read off from the integral representations, where both arise from endpoint mechanisms in the dual integration domain. It is instructive to see how these non-analyticities are encoded in the master series \eqref{eq: F++ master series}, in which every individual term is analytic at the singular points: the $\lambda_{u,v}$ dependence is polynomial, and the kernels \eqref{eq: generalised dispersive integral} are evaluated at the regular argument $\tfrac{1-\lambda_T}{2}$. Both singularities emerge as collective effects of the towers, and each sits precisely on a boundary of the convergence domain of the series, as it should.
	
	\subsection{Total-energy singularity}
	
	On the total-energy surface, the sum over $n$ converges for $|1+\lambda_T|>1$, a domain whose boundary passes precisely through $\lambda_T=0$. The mechanism is based on three facts. First, the coefficients in \eqref{eq: Q large argument} have a slowly decaying tail,
	\begin{equation}\label{eq: an tail}
		a_n=\frac{\kappa_\mu}{n}\left[1+\mathcal{O}\!\left(\tfrac{1}{n}\right)\right],
		\qquad
		\kappa_\mu=\frac{\Gamma(1+i\mu)}{2^{\frac12-i\mu}\sqrt{\pi}\,\Gamma\!\left(\tfrac12+i\mu\right)}\,.
	\end{equation}
	Second, the kernels obey the relation $\partial_{\lambda_T}T_\rho=-\rho\,T_{\rho+1}$, inherited from their definition \eqref{eq: generalised dispersive integral}. Third, at large order the integral defining $T_\rho$ is dominated by the endpoint $\lambda=1$, where the Legendre weight trivialises, $P_{i\mu-1/2}(1)=1$, so that $\rho\,T_{\rho+1}(\lambda_T)=(1+\lambda_T)^{-\rho}\left[1+\mathcal{O}(1/\rho)\right]$. Acting with $\partial_{\lambda_T}$ on \eqref{eq: F++ master series} and retaining these tails, the sum over $n$ at fixed $m$ resums into a logarithm, $\sum_{n\geq1}x^n/n=-\log(1-x)$ with $x=(1+\lambda_T)^{-2}\to1$, while the geometric sums over $m$ reconstruct the rational factors of the integrand at the endpoint,
	\begin{equation}\label{eq: m geometric}
		\sum_{m=0}^{\infty}\frac{\lambda_u^m+\lambda_v^m}{(1+\lambda_T)^{m}}
		=(1+\lambda_T)\left[\frac{1}{1+\lambda_T-\lambda_u}+\frac{1}{1+\lambda_T-\lambda_v}\right]
		\;\xrightarrow[\;\lambda_v\to-\lambda_u\;]{}\;\frac{2}{1-\lambda_u^2}\,,
	\end{equation}
	these sums being convergent for $|\lambda_{u,v}|<|1+\lambda_T|$, as holds on the total-energy surface for $|\lambda_u|<1$. Finally, all the $\mu$-dependent prefactors of \eqref{eq: F++ master series} cancel, as one has
	$\frac{\Gamma\!\left(\tfrac12+i\mu\right)}{\Gamma(1+i\mu)}\;2^{-\frac12-i\mu}\,\kappa_\mu=\frac{1}{2\sqrt{\pi}}$.
	Assembling all pieces yields $\partial_{\lambda_T}\big(s^2\,\Re\tilde{\mathcal{F}}_{++}\big)=\frac{2}{1-\lambda_u^2}\log\lambda_T+\mathcal{O}(1)$, which integrates to \eqref{eq: total energy expansion}. The subleading $\mathcal{O}(1/n)$ corrections in the tails only generate $\sum_n x^n/n^2$-type sums, which are continuous at $\lambda_T=0$ and only contribute to the analytic terms. 
	
	\subsection{Partial-energy singularity}
	
	At $\lambda_u\to-1$ with $\lambda_v>1$ fixed, it is now the sum over $m$ in the master series \eqref{eq: F++ master series} whose convergence boundary, $|\lambda_v|=|1+\lambda_T|$, passes precisely through the singular point $\lambda_u=-1$. Using the same large-order asymptotics as above, now with $\rho\to\infty$ through $m$, the tail of the $\lambda_v^m$ terms reads
	\begin{equation}
		\lambda_v^m\,T_{2n+m+\frac32+i\mu}(\lambda_T)
		\simeq \frac{1}{m}\left(\frac{\lambda_v}{1+\lambda_T}\right)^{\!m}(1+\lambda_T)^{-2n-\frac12-i\mu}\,,
	\end{equation}
	and the sum over $m$ resums into a logarithm, $\sum_{m\geq1}x^m/m=-\log(1-x)$ with $1-x=\frac{1+\lambda_T-\lambda_v}{1+\lambda_T}=\frac{1+\lambda_u}{1+\lambda_T}$, producing $-\log(1+\lambda_u)$ up to regular terms; the $\lambda_u^m$ terms, of ratio $|\lambda_u|/(1+\lambda_T)\to1/\lambda_v<1$, stay regular. The remaining sum over $n$ is then nothing but the defining series \eqref{eq: Q large argument} of the Legendre $Q$ function evaluated at $z=1+\lambda_T$: the series reassembles its own kernel, and one finds
	\begin{equation}
		s^2\,\Re\tilde{\mathcal{F}}_{++}\;\xrightarrow[\;\lambda_u\to-1\;]{}\;
		-2\,\Re \,Q_{i\mu-1/2}(1+\lambda_T)\,\log(1+\lambda_u)+\mathrm{regular}\,,
	\end{equation}
	in agreement with \eqref{eq: partial energy expansion} since $1+\lambda_T\to\lambda_v$. As with the integral representation in the main text, the similarity and difference with the total-energy case are manifest: there, the $n$ tail produced the logarithm while the geometric sums over $m$ reconstructed the rational, flat-space factor at the endpoint; here, the $m$ tail produces the logarithm while the sum over $n$ reconstructs the Legendre kernel, i.e.~the massive sub-correlator.
	
	\section{Diagrammatic rules for wavefunction coefficients}
	\label{app:WFU}
	
	Cosmological observables are directly related to primordial correlation functions. It can nonetheless be interesting to consider more primitive objects from which correlators can be deduced, namely wavefunction coefficients (see e.g.~\cite{Anninos:2014lwa,Goon:2018fyu,Goodhew:2020hob}). The latter enter into the expansion of the wavefunction for small fluctuations:
	\begin{equation}
		\Psi(\varphi,\tau_0)=\exp\left[-\sum_{n\geq2}\frac{1}{n!}\int\prod_{i=1}^n \left(\frac{\d^d\mathbf{k}_i}{(2 \pi)^d}\,\varphi_{\mathbf{k}_i}\right)\,\delta\left(\sum_{i=1}^n \mathbf{k}_i \right) \psi_n(\mathbf{k}_i)\right]\,.
		\label{wavefunction_coeffs_def}
	\end{equation}
	The wavefunction coefficients $\psi_n$ can be expressed as a sum over Feynman diagrams with essentially the same rules as for cosmological correlators, with one difference: as fields are only propagated towards the future, vertices are not decorated by Schwinger-Keldysh indices, and there is only one type of bulk-to-bulk propagator. The latter is given by
	\begin{equation}
		\begin{aligned}
			G(\tau,\tau';s)=i &\Biggl( \Theta(\tau-\tau') \sigma(\tau,s) \sigma^*(\tau',s) +  \Theta(\tau'-\tau) \sigma(\tau',s) \sigma^* (\tau,s)
			\\
			&- \frac{\sigma(\tau_0,s)}{\sigma^*(\tau_0,s)} \sigma^* (\tau,s) \sigma^*(\tau',s) \Biggr)\,,
		\end{aligned}
		\label{bulktobulk_wavefct}
	\end{equation}
	where the last, unordered, piece is chosen so that the bulk-to-bulk propagator vanishes on the late-time boundary $G(\tau_i,\tau_0,s_{ij})=0$. As for the bulk-to-boundary propagator, in this context, it is normalised to $1$ as $\tau \to \tau_0$ and $\tau_0$ is pushed to $0^{-}$, and is given by $K(E;\tau)=\frac{\tau}{\tau_0}\,e^{i E\tau}$ for a conformally coupled field. It is therefore immediate that the construction in the main text goes through, simply by taking into account these modifications. In particular, the equivalent of \eqref{eq:decomposition-ordered-lambda} for wavefunction coefficients reads
	\begin{equation}
		\begin{tikzpicture}[baseline={(0,0)}]
			\filldraw (0,0) circle (2pt) node[above] {$E_i$} node[anchor=north west] {$\lambda_{ij}$};
			\filldraw (2,0) circle (2pt) node[above] {$E_j$} node[anchor=north east] {$\lambda_{ji}$};
			\draw[thick] (0,0) to (2,0);
			\node at (1,0.3) {$s_{ij}$};
		\end{tikzpicture}=
		\begin{tikzpicture}[baseline={(0,0)}]
			\tikzset{thick curved arrow/.style={
					thick,
					decoration={markings, mark=at position 0.55 with {\arrow[scale=1.5]{latex'}}},
					postaction={decorate}
			}}
			\filldraw (0,0) circle (2pt) node[above] {$E_i$} node[anchor=north west] {$\lambda_{ij}$};
			\filldraw (2,0) circle (2pt) node[above] {$E_j$} node[anchor=north east] {$\lambda_{ji}$};
			\draw[thick curved arrow] (0,0) to (2,0);
			\node at (1,0.3) {$s_{ij}$};
		\end{tikzpicture}
		+
		\begin{tikzpicture}[baseline={(0,0)}]
			\tikzset{thick curved arrow/.style={
					thick,
					decoration={markings, mark=at position 0.55 with {\arrow[scale=1.5]{latex'}}},
					postaction={decorate}}}
			\filldraw (0,0) circle (2pt) node[above] {$E_i$} node[anchor=north west] {$\lambda_{ij}$};
			\filldraw (2,0) circle (2pt) node[above] {$E_j$} node[anchor=north east] {$\lambda_{ji}$};
			\draw[thick curved arrow] (2,0) to (0,0);
			\node at (1,0.3) {$s_{ij}$};
		\end{tikzpicture}
		+
		\begin{tikzpicture}[baseline={(0,0)}]
			\tikzset{thick curved arrow/.style={
					thick,
					decoration={markings, mark=at position 0.55 with {\arrow[scale=1.5]{latex'}}},
					postaction={decorate}}}
			\filldraw (0,0) circle (2pt) node[above] {$E_i$} node[anchor=north west] {$\lambda_{ij}$};
			\filldraw (2,0) circle (2pt) node[above] {$E_j$} node[anchor=north east] {$\lambda_{ji}$};
			\draw[dashed] (2,0) to (0,0);
			\node at (1,0.3) {$s_{ij}$};
		\end{tikzpicture}
		\label{eq:decomposition-ordered-WFU}
	\end{equation}
	where the first two terms are the same as for correlators, given in \eqref{integral_single_exch_1_2}-\eqref{integral_single_exch_2_1}, and the last term contributes as  
	\begin{equation}
		\begin{tikzpicture}[baseline={(0,0)}]
			\tikzset{thick curved arrow/.style={
					thick,
					decoration={markings, mark=at position 0.55 with {\arrow[scale=1.5]{latex'}}},
					postaction={decorate}}}
			\filldraw (0,0) circle (2pt) node[above] {$E_i$} node[anchor=north west] {$\lambda_{ij}$};
			\filldraw (2,0) circle (2pt) node[above] {$E_j$} node[anchor=north east] {$\lambda_{ji}$};
			\draw[dashed] (2,0) to (0,0);
			\node at (1,0.3) {$s_{ij}$};
		\end{tikzpicture}
		=\int_{-\infty}^0 \d\tau_i \int_{-\infty}^0 \d\tau_j \, e^{i E_i \tau_i} e^{i E_j \tau_j} \,e^{i s_{ij} \lambda_{ij}(1-i \epsilon) \tau_i} e^{i s_{ij} \lambda_{ji}(1+i \epsilon) \tau_j}
		\,.
		\label{eq:new-term-WFU}
	\end{equation}
	Here, we considered complementary series fields for which
	\begin{equation}
		\frac{\sigma(\tau_0,s)}{\sigma^*(\tau_0,s)} e^{-i\pi\left(i \mu+\frac{1}{2}\right)}\underset{\tau_0\rightarrow 0}{\rightarrow}1\,,
	\end{equation}
	which simplifies the contribution from the last term in \eqref{bulktobulk_wavefct}. It is easy to track the corresponding factor multiplying \eqref{eq:new-term-WFU} for principal series fields, but it is not needed: cosmological correlators are analytic functions of the exchanged masses, hence it is sufficient to proceed with \eqref{eq:decomposition-ordered-WFU}, and simply analytically continue the total physical result.

	For convenience, as in Section~\ref{sec:rules-Laplace}, we summarise the complete set of Laplace-space diagrammatic rules that allow us to write directly any diagram contributing to wavefunction coefficients in terms of Laplace-space integrals.
	
	\begin{enumerate}
		\item Label each vertex of the diagram under consideration with a number $i$ and compute the number
		\begin{equation}
			N_i=C_i-4+2 M_i\,,
			\label{power_exponent_vertex_summary}
		\end{equation}
		where $C_i$ is the number of external conformally coupled legs and $M_i$ is the number of internal massive lines connected to $i$. We always have $N_i\geq 0$ for physically relevant diagrams. Then, apply the differential operator
		\begin{equation}
			\left(- \,i\right)^{N_i}\frac{\partial}{\partial E_i^{N_i}}\,,
			\label{diff_op_Ei_summary}
		\end{equation}
		at the end of the computation. If there is no external leg connected to the vertex $i$, introduce the vertex energy $E_i>0$ as a regulator (without modification of the number $N_i$), apply the previous differential operator and take the limit $E_i\rightarrow 0$.
		\item Assign the dual variable $\lambda_{ij}$ to the vertex $i$ and the internal line connecting vertices $i$ and $j$. Decompose each internal line as
		\begin{equation}
			\begin{tikzpicture}[baseline={(0,0)}]
				\filldraw (0,0) circle (2pt) node[above] {$E_i$} node[anchor=north west] {$\lambda_{ij}$};
				\filldraw (2,0) circle (2pt) node[above] {$E_j$} node[anchor=north east] {$\lambda_{ji}$};
				\draw[thick] (0,0) to (2,0);
				\node at (1,0.3) {$s_{ij}$};
			\end{tikzpicture}=
			\begin{tikzpicture}[baseline={(0,0)}]
				\tikzset{thick curved arrow/.style={
						thick,
						decoration={markings, mark=at position 0.55 with {\arrow[scale=1.5]{latex'}}},
						postaction={decorate}
				}}
				\filldraw (0,0) circle (2pt) node[above] {$E_i$} node[anchor=north west] {$\lambda_{ij}$};
				\filldraw (2,0) circle (2pt) node[above] {$E_j$} node[anchor=north east] {$\lambda_{ji}$};
				\draw[thick curved arrow] (0,0) to (2,0);
				\node at (1,0.3) {$s_{ij}$};
			\end{tikzpicture}
			+
			\begin{tikzpicture}[baseline={(0,0)}]
				\tikzset{thick curved arrow/.style={
						thick,
						decoration={markings, mark=at position 0.55 with {\arrow[scale=1.5]{latex'}}},
						postaction={decorate}}}
				\filldraw (0,0) circle (2pt) node[above] {$E_i$} node[anchor=north west] {$\lambda_{ij}$};
				\filldraw (2,0) circle (2pt) node[above] {$E_j$} node[anchor=north east] {$\lambda_{ji}$};
				\draw[thick curved arrow] (2,0) to (0,0);
				\node at (1,0.3) {$s_{ij}$};
			\end{tikzpicture}
			+
			\begin{tikzpicture}[baseline={(0,0)}]
				\tikzset{thick curved arrow/.style={
						thick,
						decoration={markings, mark=at position 0.55 with {\arrow[scale=1.5]{latex'}}},
						postaction={decorate}}}
				\filldraw (0,0) circle (2pt) node[above] {$E_i$} node[anchor=north west] {$\lambda_{ij}$};
				\filldraw (2,0) circle (2pt) node[above] {$E_j$} node[anchor=north east] {$\lambda_{ji}$};
				\draw[dashed] (2,0) to (0,0);
				\node at (1,0.3) {$s_{ij}$};
			\end{tikzpicture}
			\label{eq:decomposition-ordered-WFU-summary-rules}
		\end{equation}
		A diagram that has $m$ internal lines therefore decomposes into $3^{m}$ ordered contributions. We also define time-lines, which are sequences of internal lines obtained by following arrows. When considering a loop diagram, contributions involving closed time-lines vanish.
		\item For each variable $\lambda_{ij}$, write the following integral and integration measure (a diagram that has $m$ internal lines is then expressed in terms of $2 m$ $\lambda$-integrals):
		\begin{equation}
			\int_1^\infty \d \lambda_{ij}\,P_{i\mu_{ij}-1/2}(\lambda_{ij})\,,
		\end{equation}
		where $\mu_{ij}$ is the mass parameter of the field exchanged between vertices $i$ and $j$.
		\item For each contribution, compute the $\lambda$-space integrand:
		\begin{itemize}
			\item For a vertex $i$, compute the quantity $V_i$ that is the sum of the following contributions:
			\begin{itemize}
				\item add the vertex energy $+E_i$;
				\item for an outgoing arrow or an unordered line related to a vertex $j$, add a $+s_{ij}\,\lambda_{ij}$ factor;
				\item for an ingoing arrow related to a vertex $j$, add a $-s_{ij}\,\lambda_{ij}$ factor.
			\end{itemize}
			\item Compute $\mathcal{V}_i =i (V_i +\sum_j V_j)+\epsilon$ where the sum is over earlier vertices (i.e.\ vertices related to $i$ by ingoing arrows in a time-line).
			\item If two or more time-lines are crossing without ending at a vertex, sum over all possible time orderings of later vertices. For a given ordering, add to the computation of $\mathcal{V}_i$ contributions $V_j$ from earlier vertices according to this ordering.
			\item The $\lambda$-space integrand is given by the product over vertices $\prod_i \frac{1}{\mathcal{V}_i}$.
		\end{itemize}
		\item The Laplace-space expression of the diagram is then proportional to
		\begin{equation}
			\begin{aligned}
				\int_1^{+\infty} \left(\prod_{i,j\in\mathcal{V},i\neq j} \d\lambda_{ij} \,P_{i \mu_{ij} -\frac{1}{2}}(\lambda_{ij} )\right)
				\left(\prod_{i\in\mathcal{V}} \left(- \,i\right)^{N_i}\frac{\partial}{\partial E_i^{N_i}}
				\;\sum_{\text{time ordering}} \frac{1}{\mathcal{V}_i(E_i,s_{ij},\lambda_{ij})}\right)
			\end{aligned}
			\label{general_lambda_diagram-WFU}
		\end{equation}
		where $\mathcal{V}$ is the set of vertices and the sum is over all possibilities in the decomposition \eqref{eq:decomposition-ordered-WFU-summary-rules} for each internal line. For a loop diagram, add loop momenta integrals, and modify the first parenthesis to incorporate dual variables for each massive mode propagating in each loop internal line. 
	\end{enumerate}

	\bibliographystyle{JHEP}
	\bibliography{references-Laplace-clean.bib}
\end{document}